\documentclass[review,3p]{elsarticle}
\usepackage{overpic}
\usepackage{subfigure}
\usepackage{comment}
\usepackage{bm}
\usepackage{graphicx}
\usepackage{wrapfig}
\usepackage{booktabs}
\usepackage[colorlinks,linkcolor=blue]{hyperref}
\UseRawInputEncoding
\biboptions{numbers,sort&compress}
 \usepackage{mathrsfs}
\usepackage{makecell}
\usepackage{mathrsfs}
\usepackage{setspace}
\usepackage{multirow} %multirow for format of table
\usepackage{xcolor}
\usepackage{amssymb}
\usepackage{amsthm}
\usepackage{amsmath}
\usepackage{bm}
\usepackage{hyperref}

\usepackage{tabularx}
\usepackage{booktabs}
\usepackage{multirow}
\usepackage{makecell}

\usepackage{algorithm}
\usepackage{algpseudocode}

\usepackage{lineno}

\graphicspath{{./fig/}}
\newcommand{\tr}{\textcolor[rgb]{1.0,0.0,0.0}}

\def \mD {\mathcal{D}}

\def \mB {\mathcal{B}}
\def \mS {\mathcal{S}}

\allowdisplaybreaks[4]

\begin{document}
\begin{frontmatter}

\title{Constructing Boundary-identical Microstructures via Guided Diffusion for Fast Multiscale Topology Optimization}

\author[zero]{Jingxuan Feng\fnref{equal1}} \ead{fjx1376216197@mail.ustc.edu.cn}
\author[zero]{Lili Wang\fnref{equal1}} \ead{llw91@mail.ustc.edu.cn} \fntext[equal1]{These authors contributed equally to this work.}
\author[zero]{Xiaoya Zhai\corref{cor}} \ead{xiaoyazhai@ustc.edu.cn} \cortext[cor]{Corresponding author}
\author[first]{Kai Chen} %\ead{chenkai.cn@hotmail.com}
\author[second]{Wenming Wu} %\ead{wwming@hfut.edu.cn}
\author[zero]{Ligang Liu} %\ead{lgliu@ustc.edu.cn}
\author[zero]{Xiao-Ming Fu\corref{cor}} \ead{fuxm@ustc.edu.edu}

\address[zero]{School of Mathematical Sciences, University of Science and Technology of China, Hefei, 230026, People's Republic of China}
\address[first]{Beijing Academy of Artificial Intelligence, Beijing 100190, People's Republic of China} 
\address[second]{Hefei University of Technology, Hefei, Anhui 230026, People's Republic of China}

\begin{abstract}
Hierarchical structures exhibit critical features across multiple scales. However, designing multiscale structures demands significant computational resources, and ensuring connectivity between microstructures remains a key challenge. 
To address these issues, \textit{\textbf{large-range, boundary-identical microstructure datasets}} are successfully constructed, where the microstructures share the same boundaries and exhibit a wide range of elastic moduli. 
This approach enables highly efficient multiscale topology optimization. 
Central to our technique adopts a deep generative model, guided diffusion, to generate microstructures under the two conditions, including the specified boundary and homogenized elastic tensor. 
%active leanring, 不断的扩充数据集, 使得model的效果进一步提升
We generate the desired datasets using active learning approaches, where microstructures with diverse elastic moduli are iteratively added to the dataset, which is then retrained.  %through a deep generative model
%We achieve the desired datasets by active learning approaches which are alternately adding microstructures with diverse elastic modulus constructed by the deep generative model into the dataset and retraining the deep generative model.
%实现了几个覆盖范围广的数据集，并且数据集中有接近理论极限的结构。
After that, sixteen boundary-identical microstructure datasets with wide ranges of elastic modulus %high property coverage 
are constructed.  %by our method.
%进行了multiscale design例子，达到了什么结果
We demonstrate the effectiveness and practicability of the obtained datasets over various multiscale design examples. 
Specifically, in the design of a mechanical cloak, we utilize macrostructures with $30 \times 30$ elements and microstructures filled with $256 \times 256$ elements. The entire reverse design process is completed within one minute, significantly enhancing the efficiency of the multiscale topology optimization.
%In the design of the mechanical cloak, we utilize macrostructures with $30 \times 30$ elements and fill microstructures with $256 \times 256$ elements. The entire reverse design process is completed within one minute, significantly enhancing the efficiency of multiscale optimization.
\end{abstract}


\end{frontmatter}

\section{Introduction} \label{sec:Intro}

Microstructures are characterized by their unique physical properties derived from specialized topological structures. %, constitute exceptional artificial materials. 
Heterogeneous multiscale systems composed of microstructures can achieve sophisticated functionalities, such as invisibility cloaks~\cite{optical,cloaked}, soft robotics~\cite{kirigami}, and high thermal conductivity~\cite{feng2024topology, dirker2013topology}. 
Despite advances in modern manufacturing technologies that enable the fabrication of these intricate systems spanning macroscopic and microscopic scales, designing multiscale systems remains a challenge, primarily involving intricate inverse design of microstructures and costly nested multiscale optimization processes.

\begin{figure}[t]
    \centering
    \includegraphics[width=0.8\linewidth]{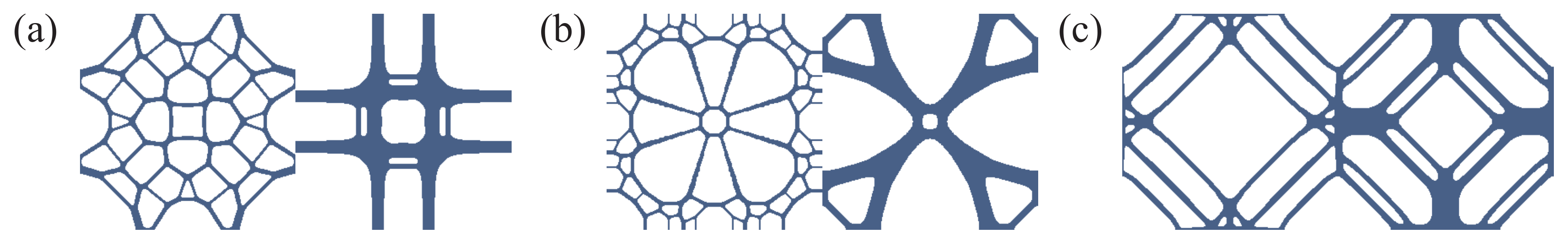}
    \caption{Illustration of boundary connections. (a) The boundary is completely disconnected. (b) The boundary is partly connected.
    (c) The boundary is fully connected, known as boundary-identical microstructure sets.}
    \label{fig:boundarycond}
\end{figure}

Advances in computational capabilities and machine learning have spurred interest in data-driven multiscale optimization. 
The existing frameworks are broadly categorized into top-down methods~\cite{td0,td1,td2} and bottom-up approaches~\cite{bu0,bu1,bu2}. 
Top-down methods first perform topology optimization to determine macroscopic property distributions, followed by populating the design with microstructures from a pre-generated database. 
Bottom-up approaches use microstructural parameters, such as volume fraction, as design variables, using machine learning models to predict effective properties and accelerate optimization. 
However, the parameterized representation in bottom-up approaches may limit the property space of microstructures. Thus, we adopt a top-down approach, which relies on a comprehensive microstructure database as the building blocks. The design process involves matching microstructures to desired properties while ensuring two key requirements: (1) Boundary Connectivity: Essential for periodic or graded arrangements, affecting the transmission of forces, waves, etc. Figure~\ref{fig:boundarycond} categorizes connections as unconnected, partially connected, or fully connected (identical boundary), with compatible microstructures offering improved performance~\cite{GARNER201965}. (2) Elastic Moduli Range: A large-range dataset covering a wide range of elastic moduli enhances multiscale design by expanding the feasible design space and increasing alignment with required properties.

However, generating such a microstructure dataset is challenging. The reasons are twofold. 
First, the diversity of generated structures of previous structural optimization algorithms~\cite{wang2016parametric,zobaer2020energy,zhao2023study,10.1145/2766926, cramer2016microstructure, wang2017concurrent,GARNER201965,ZHAI2024116530,Du2018} is poor. Moreover, the hard connectivity constraint further limits the range of elastic moduli. %property coverage.
Second, choosing a boundary for all microstructures to realize the goal of a wide range of elastic moduli %high property coverage 
is extremely difficult or clueless.  

Three primary methods for generating compatible microstructures are parametric modeling~\cite{wang2016parametric,zobaer2020energy,zhao2023study}, geometric interpolation~\cite{10.1145/2766926,cramer2016microstructure,wang2017concurrent}, and connectivity topology optimization~\cite{GARNER201965,ZHAI2024116530,Du2018}. 
Parametric modeling adjusts predefined microstructures to meet property requirements through an iterative trial-and-error process. 
Geometric interpolation generates a microstructure sequence by interpolating between two predefined deformed structures. 
Connectivity topology optimization enforces connectivity constraints using gradient descent to achieve geometric continuity. %But, prior methods failed to ensure fully identical boundaries. ncomplete boundary connectivity can
However, they fail to ensure fully identical boundaries and may cause faults, stress concentrations, and manufacturing challenges. 
Furthermore, none of these methods effectively generate large-range microstructure datasets with a broad range of elastic moduli.

%\todo{Fifth paragraph: our method}
Identical boundary embedded large-range microstructure datasets are successfully constructed in this paper. 
%This paper proposes a novel method to construct microstructure datasets with fully connected boundaries and a wide range of elastic moduli. %high property coverage. 
Boundary connectivity constraints can be enforced using the same boundary and symmetry properties. In addition, we opt for cubic symmetric microstructures for simplicity of calculation. 
%如何实现diversity and high property coverage under the hard compatibility constraint
The key to our algorithm is using a deep generative model to achieve diverse microstructures with large elastic moduli coverage under boundary-identical and cubic symmetric constraints. 
Specifically, we adopt a self-conditioning diffusion model to generate microstructures under two conditions: (1) the predefined boundary and (2) the specified homogenized elastic tensor.

%high property coverage
%这句话是不是有重复表达的意思？
%\tr{We can achieve a wide range of elastic moduli when the coverage space of specified homogenized elastic tensors is large enough.} 
Training the diffusion model requires a dataset. However, constructing this dataset is our primary objective, creating a challenging chicken-and-egg problem. 
%如何choose a boundary for all structures
To this end, we propose a practical strategy. %constructing the desired datasets contains the following steps.
First, we train a self-conditioning diffusion model using a microstructure dataset constructed by an optimization algorithm that incorporates cubic symmetry constraints but excludes the boundary-identical condition. We also cluster the microstructure boundaries to generate a small set of seed boundaries. 
Then, we run the diffusion model for each seed boundary to generate diverse microstructures under various homogenized elastic tensors while fixing the seed boundary. 
These microstructures form a boundary-identical dataset corresponding to each seed boundary. 
%which are further used to train the self-conditioning diffusion model.
%Finally, we obtain the desired dataset by iteratively retraining a diffusion model using each boundary-identical dataset and adding the constructed microstructures using the trained diffusion model into the dataset.
Finally, we create the desired dataset by iteratively retraining a diffusion model on each boundary-identical dataset and adding the generated microstructures. 

\begin{figure*}[t]
    \centering
    \includegraphics[scale=0.08]{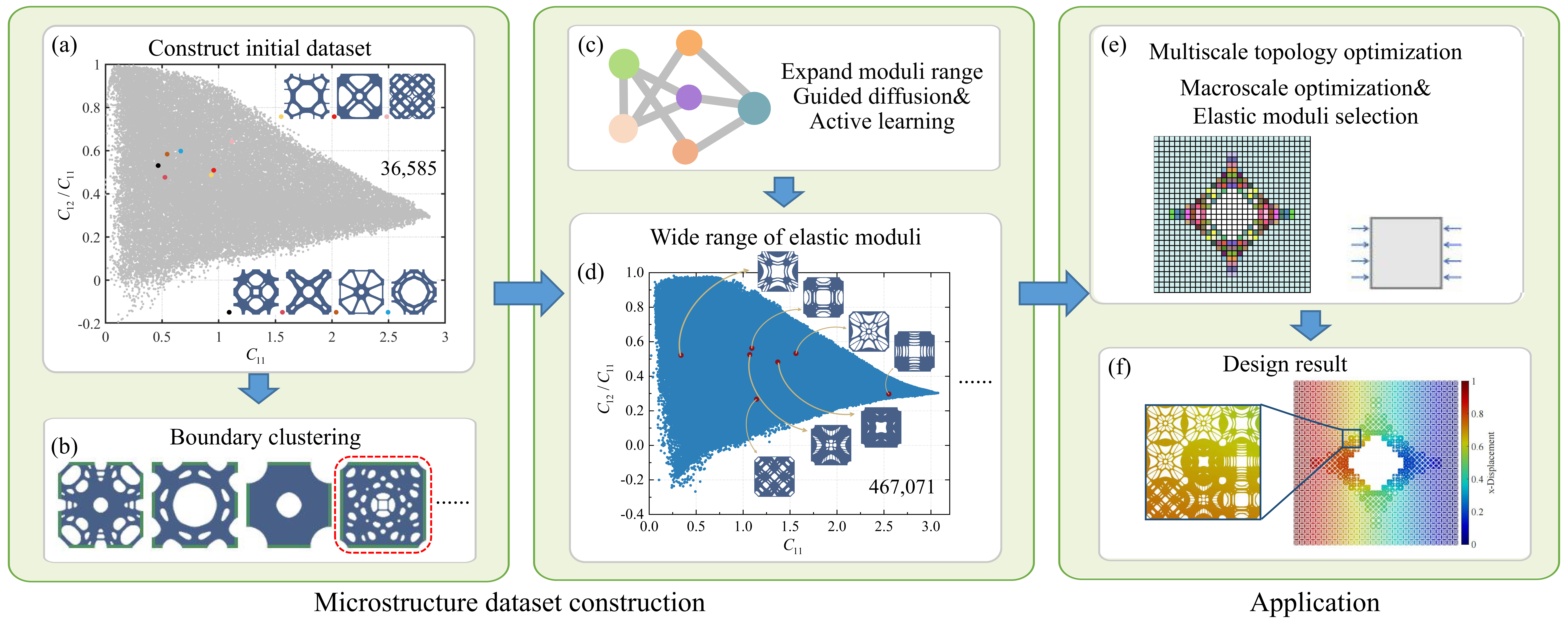}
    \caption{The pipeline of constructing large-range boundary-identical microstructure datasets and their applications.  
    (a) The cubic symmetric microstructure dataset without identical boundaries. 
    %The dataset generated by optimization and perturbation. %, uses circles to represent microstructures with arbitrary boundaries and hexagrams to represent microstructures with similar boundaries. 
    %The dataset contains a total of 36,585 microstructures. 
    (b) Representative boundaries derived from boundary clustering. 
    %Microstructural boundaries are outlined in green lines, and red dashed boxes highlight the selected boundaries that will proceed to the guided diffusion step. 
    (c) Generated microstructure datasets with the specified boundaries, which are expanded 
    through guided diffusion and active learning. 
    (d) A dataset with a wide range of elastic moduli.  %generated using a diffusion model, 
    %where all boundaries are identical and match the boundaries of the microstructure enclosed by the red dashed box in (b). 
    (e)-(f) %The Elastic moduli results derived from 
    The results of multiscale topology optimization. %, with different colors indicating microstructures having different modulus values. 
     %Using the dataset with the same boundaries as shown in (d) for assembly, the resulting multiscale design ensures perfect alignment along boundaries.
    } %at the connections of each cell. 
    %\tr{font revise...}
    %Microstructure datasets with identical boundaries (left), along with the mechanical cloak design results derived from the dataset (right).
    
    \label{fig:one-result}
\end{figure*}

%\todo{Sixth paragraph: experiments} refer to Figure~\ref{fig:one-result}. 
We are the first to construct large-range microstructure datasets with the same boundary and wide coverage of elastic moduli. %high property coverage. 
The dataset is inherently conducive to multiscale design and effectively addresses a crucial issue in multiscale structures - the connectivity problem.  
We utilize the dataset for two multiscale design scenarios: a mechanical cloak design and a customized displacement design. By testing various boundary cases and hole configurations, we demonstrate the feasibility and versatility of the datasets. It is noteworthy that in the design of the mechanical cloak, %we employ $30\times30$ macro units and fill microstructures with $256\times256$ elements. The entire reverse design process is accomplished within one minute, markedly enhancing the efficiency of multiscale optimization.
we employ $30\times30$ macro units and microstructures filled with $256\times256$ elements. The entire reverse design process is accomplished within one minute, markedly enhancing the efficiency of multiscale optimization. Our pipeline is shown in Figure~\ref{fig:one-result}. %(Figure~\ref{fig:one-result}).
Code for this paper is available at \url{https://github.com/Schnabel-8/BDR-diffusion}, and the datasets are publicly available at %\url{https://rec.ustc.edu.cn/share/b9e072b0-306a-11ef-91cb-dfb6325d6cc5}.
\href{https://rec.ustc.edu.cn/share/b9e072b0-306a-11ef-91cb-dfb6325d6cc5}{https://rec.ustc.edu.cn/share/dataset}. 
In summary, our technical contributions are as follows:
\begin{itemize}
    \item Under the boundary-identical constraint, large-range microstructure datasets are successfully constructed. \textbf{This is proved to be feasible for the first time.}
    \item A fast multiscale optimization strategy using large-range, boundary-identical microstructure datasets achieves significant acceleration over state-of-the-art methods. %A fast multiscale structure optimization strategy is developed with generated microstructure datasets as building blocks. \textbf{Compared with the state-of-the-art methods, we achieve an acceleration of about four orders of magnitude under the resolution of $50\times 50$.}
\end{itemize}

%---------------------------------------------
\vspace{-0.5cm}
\section{Related works} \label{sec:rel}
\paragraph{Microstructure connectivity constraints} 
Three methods \text{-} parametric modeling~\cite{wang2016parametric, zobaer2020energy, zhao2023study}, geometric interpolation~\cite{10.1145/2766926, cramer2016microstructure, wang2017concurrent}, and topology optimization~\cite{GARNER201965, ZHAI2024116530, Du2018} - play crucial roles in generating microstructures with good connectivity. 
Parametric modeling effectively describes microstructure geometries for design purposes, yet achieving structures meeting specific performance criteria involves a trial-and-error process. 
Similarly, geometric interpolation ensures microstructural connectivity, but assessing its performance improvement remains challenging. 
Consequently, they are usually integrated with topology optimization techniques to achieve improved results. 
Zobazer and Sutradhar~\cite{zobaer2020energy} introduce supershape-based parametric modeling~\cite{gielis2003generic} to ensure smooth connectivity at interfaces. %They utilize supershape parameters to model voids, adjusting these parameters to generate connect microstructures. 
Schumacher et al.~\cite{10.1145/2766926} achieve connected microstructures by interpolating from a microstructure database. 
Shape interpolation methods are developed using signed distance functions or characteristic level set functions~\cite{cramer2016microstructure, wang2017concurrent}. 
Du et al.~\cite{Du2018} combine topology optimization with newly defined connectivity index constraints. 
Garner et al.~\cite{GARNER201965} consider geometric connectivity and physical compatibility for the inverse homogenization method. 
Zhai et al.~\cite{ZHAI2024116530} propose a differentiable microstructure generation framework, reformulating the problem as a non-uniform heat diffusion process. 
Despite these advancements, they still need to ensure complete boundary connectivity. 
Moreover, the initial microstructure will influence the final design, leading to different structures and properties with varied initializations. 
While all three methods ensure microstructural connectivity, they impose varying limitations on the diversity of the compatibility dataset. 
Instead, we propose constructing cubic symmetric microstructures with identical boundaries and diverse geometries to ensure complete boundary connectivity.

\paragraph{Microstructure dataset} 
Microstructure datasets encompass geometric configurations, physical properties, and performance characteristics. 
They serve as essential tools for analysis and prediction goals~\cite{xu2021method,doi:10.1126/sciadv.aao7005,wang2021data,peng2022ph}, experimentation/simulation~\cite{steinmetz2016analytics,pilchak2016dataset,yang2018deep,wang2022mechanical}, and the development of models or algorithms in microstructure design~\cite{wang2020deep,rade2021algorithmically,ahmad2022ann,patel2022improving,seo2023development}. 
Currently, microstructure design-oriented datasets consist primarily of density-based 2D and 3D datasets~\cite{wang2020data, chan2021metaset,10.1115/1.4055925}, along with truss-based datasets~\cite{korshunova2021uncertainty,bastek2022inverting,zheng2023unifying}. The construction of microstructure datasets is often based on tasks, where factors such as dataset size and property coverage can influence the effectiveness of downstream tasks. 
The methods employed for dataset generation can be broadly categorized into three main types: heuristic generation~\cite{chan2020metaset,chan2021metaset,wang2023data,zheng2023unifying}, perturbation generation~\cite{lissner2019data,wang2020data,robertson2023local,lambard2023generation}, and simulation generation~\cite{steinmetz2016analytics,yang2018deep,li2021towards,wang2022ih}. 
Shape-driven heuristics quickly generate data but lack control, leading to unpredictable and low-quality datasets. Effective property calculation demands costly finite element simulations, making high-quality dataset acquisition challenging. For multiscale design, datasets must have broad property coverage and diverse geometries to ensure compatibility during the filling process.

\paragraph{AI for microstructures}
Neural networks have been successfully used for generating a variety of microstructures (c.f. the survey in~\cite{lee2023data}). 
%A comprehensive review is given in Paper~\cite{lee2023data}. 
Examples include obtaining the homogenization of composite structures~\cite{le2015computational,liu2021review,peng2022ph}, effective response of stress-strain curves~\cite{yuan2018machine,long2021modeling,wu2023developing,zheng2023unifying}, and extreme mechanical microstructures~\cite{wu2020machine,challapalli2023inverse,doi:10.1126/sciadv.aao7005}. %\tr{Different methods discussion...}
%AI methods have been successfully employed in designing high-performance microstructures. 
In terms of performance prediction, AI can establish predictive models by learning from a vast amount of microstructure data and their corresponding performance characteristics~\cite{kautz2021predicting,10.1115/1.4055925,peng2022ph}. 
This enables researchers to rapidly screen potential exceptional microstructure candidates and save significant time in experimentation and simulation. 
For example, Le et al.~\cite{le2015computational} propose a decoupled computation homogenization method for nonlinear elastic materials using neural networks. 
This method utilizes a neural network model to compute the effective potential. 
Deng et al.~\cite{deng2022inverse} design a network to predict the nonlinear response curves of hinged quadrilateral microstructures. %Using this network, they performed inverse design to create energy-absorbing structures and soft robots. 
Ma et al.~\cite{ma2022deep} use residual networks instead of finite element analysis to compute strains and successfully employ genetic algorithms for the inverse design of microstructures with predetermined global strains under magnetic actuation. 
On the other hand, in the process of inverse design, AI models, particularly generative models, can effectively explore complex design spaces and discover new structures and material combinations that go beyond the limitations of human intuition~\cite{chen2018computational}. 
Zheng et al.~\cite{zheng2023unifying} utilize a graph-based VAE to design truss structures with customized mechanical properties in both linear and nonlinear states, including designs exhibiting exceptionally stiff, auxetic, pentamode-like, and tailored nonlinear behaviors. 
Wang et al.~\cite{wang2022ih} propose a method using inverse homogenization generative adversarial networks (IH-GANs) for designing variable-density cellular structures. 
Zheng et al.~\cite{zheng2021controllable} develop a GAN-based design method for auxetic microstructures and successfully use this method to design auxetic microstructures with predetermined Young's modulus and Poisson's ratio. 
Diffusion models are currently the mainstream approach for content generation. According to \cite{dhariwal2021diffusion}, diffusion models surpass previous methods such as GAN and VAE in terms of the quality of generated contents. Consequently, we employ conditional diffusion models for the inverse generation of microstructures.
%VAEs face challenges in generating high-quality results, often necessitating post-processing. GANs suffer from training difficulties and instability. 
%In contrast, based on a diffusion model, our method offers enhanced stability during training and significantly improved generated quality compared to previous models.
%However, VAEs have limitations in terms of generated quality, and the generated results often require post-processing. The biggest issue with GANs is their difficulty in training and lack of stability. Our method is based on diffusion model, which is more stable during training and has significantly improved generated quality compared to previous models.

%-------------------------------------------------
\section{Method} \label{sec:method}

\subsection{Overview}
\paragraph{Goals and requirements}
Our goal is to construct a large-range microstructure dataset $\mD$ satisfying the following requirements:
\begin{enumerate}
    \item \emph{Cubic symmetric}: each microstructure $\mathbf{s}$ is cubic symmetric.
    \item \emph{Identical boundary}: the microstructures in the dataset have an identical boundary $\mB$.
    \item \emph{Wide range of elastic moduli}: the physical properties of the microstructures cover a large space $\mS$.
\end{enumerate}
The first and second requirements are hard constraints to ensure that the microstructures can achieve full compatibility in multiscale design.
The third requirement aims to increase the choice space in multiscale design significantly.

\paragraph{Methodologies}
%\todo{diversity -> deep generative model, self-conditioning diffusion model}
To ensure the diversity of microstructures, we use a deep generative model instead of the optimization methods. 
We first %develop a practical method to 
construct a dataset $\hat{\mD}$ satisfying the first and second requirements (Section~\ref{sec:initdata}).
%The k-means clustering operation is adopted to extract boundary. %(Section~\ref{sec:initdata}). 
Specifically, to reduce the difficulty of meeting the three requirements, we use a self-conditioning diffusion model~\cite{zheng2023lasdiffusion} with property guidance, where the properties including the specified boundary $\mB$ and homogenized elastic tensor $C$. 
Finally, to realize the third requirement, we propose a pipeline based on active learning that alternately runs the following two steps (Section~\ref{sec:network}).  %(Figure~\ref{fig:pipeline}):
\begin{enumerate}
    \item \emph{Training the self-conditioning diffusion model}: using the dataset satisfying the first and second requirements to train the self-conditioning diffusion model.
    \item \emph{Augmenting the dataset}: adding the microstructures generated by the self-conditioning diffusion model into the dataset.
\end{enumerate}

\subsection{Initializing microstructure dataset} \label{sec:initdata}

\paragraph{Representation}
We discrete each microstructure $\mathbf{s}$ into $256\times 256$ elements denoted as a binary matrix $\mathcal{A}_{\mathbf{s}}$, where 0 represents void and 1 represents solid. The number of elements $256\times256$ effectively represents diverse microstructures and offers satisfactory computational efficiency. We characterize the microstructure boundaries $\mathcal{B}_{\mathbf{s}}$ through a binary vector $\mathbf{b}_\mathbf{s}$ %of length 256. Specifically, for a microstructure represented by a binary matrix, we 
extracted from the first row of the matrix $\mathcal{A}_{\mathbf{s}}$.

\begin{figure}[t]
    \centering
    \includegraphics[width=0.85\linewidth]{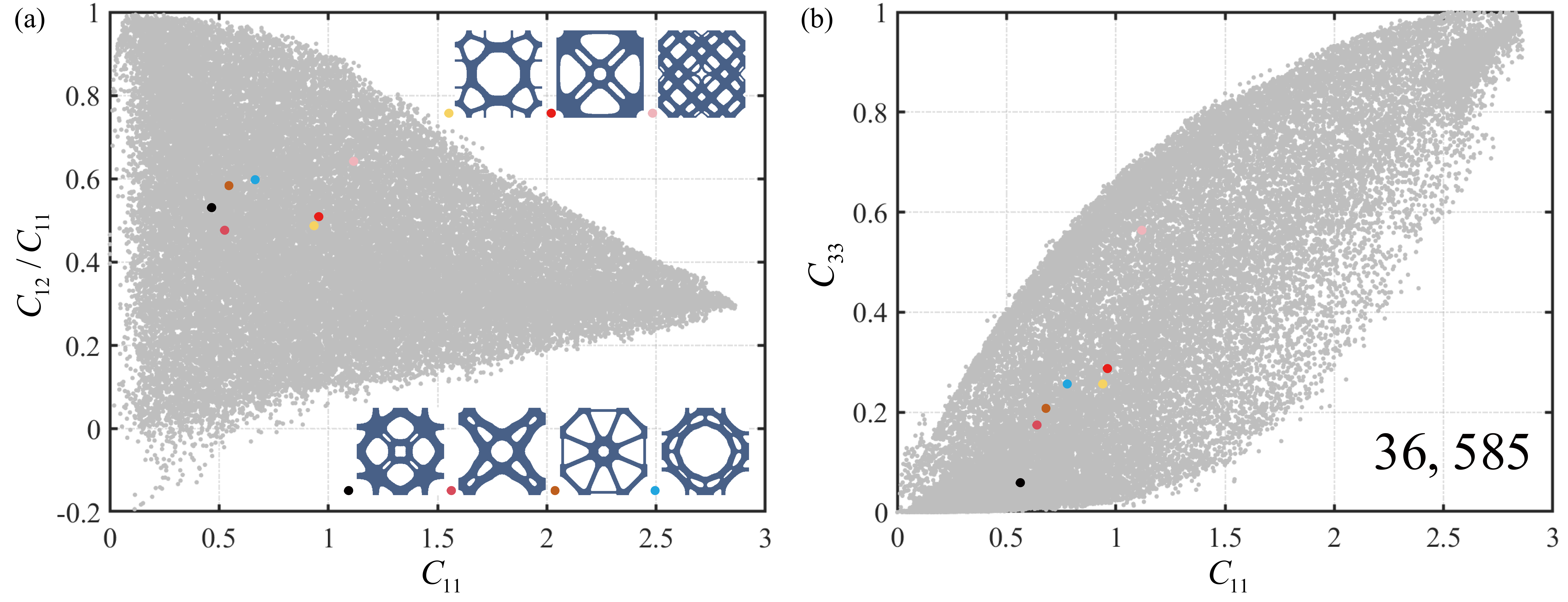}
    \caption{Plots of $C_{11}$ vs. $C_{12}/C_{11}$ (a) and $C_{11}$ vs. $C_{33}$ (b) on a cubic symmetric microstructure dataset $\mathcal{D}_0$ without identical boundaries. The number of microstructures is 36,585.
    }
    \label{fig:boundarycon}
\end{figure}

\paragraph{Dataset via optimization, perturbation and filtering}
We use the optimization method~\cite{xia2015design} %with 
and shape perturbation~\cite{kroon} to generate a cubic symmetric microstructure dataset \tr{$\mD_0$} (shown in Figure~\ref{fig:boundarycon}) without identical boundaries. 
%requirements 1 and 3:
%high property coverage:
\begin{enumerate}
    \item \emph{
    Optimizing microstructures}: we generate microstructures by optimizing the bulk and shear modulus under various volume fractions via inverse homogenization method~\cite{xia2015design}. % Specifically, the volume fraction ranges from 0.2 to 0.9, generating 16,000 distinct microstructures.
    Specifically, we consider a volume fraction range of 0.2 to 0.9 and generate 16,000 microstructures.
    %This process involves considering a
    \item \emph{Perturbing microstructures}: utilizing the radial distortion model proposed by Kroon~\cite{kroon}, each microstructure is subjected to perturbation. For each microstructure, we generate 30 different perturbed versions, resulting in a total of 480,000 perturbed microstructures. %This step generates 30 perturbed versions for each original microstructure, totaling 48,000 perturbed microstructures.
    The perturbed microstructures are filtered to retain all fully connected microstructures while discarding the disconnected ones. There are about 400,000 microstructures left.
    \item \emph{Eliminating similarity:} %\tr{Merging} %Organizing 
    %and selecting microstructures}:  
    We merge the perturbed microstructures with the optimized ones and refine the dataset to eliminate microstructures that are geometrically similar. %Afterwards, we merge the perturbed microstructures with the optimized microstructures and then remove the geometrically similar structures, and finally obtain a dataset of 36,585 microstructures. %resulting in a final dataset size of 36,585.
    Let the merged dataset be denoted as \(\bar{\mD}\). We first perform uniform sampling within the coverage range of the elastic tensors in the dataset. The sampled dataset is referred as \(\bar{\mD}'\). For each microstructure $\mathbf{s}$ in dataset \(\bar{\mD}'\), we calculate its similarity to other structures within the dataset. Structure similarity (SS) is defined as:
    
    \begin{equation} \label{eq:ss}
    \text{SS}(\mathbf{s}) = \max \left( \{ \text{IoU}(\mathcal{A}_\mathbf{s},\mathcal{A}_\mathbf{s'}) : \mathbf{s'} \in N(\mathbf{s},\bar{\mathcal{D}}') \} \right), \end{equation}
    where the set \(N(\mathbf{s},\bar{\mathcal{D}}') = \{\mathbf{t} \,| \,\mathbf{t} \in \bar{\mD}', \mathbf{t}\neq \mathbf{s} \text{ and } \|(C_{11}(\mathbf{s}), C_{12}(\mathbf{s}), C_{13}(\mathbf{s})) - (C_{11}(\mathbf{t}), C_{12}(\mathbf{t}), C_{13}(\mathbf{t}))\|_2 < 0.5\}\), where \(C_{ij}(\mathbf{s})\) (for 2D cases, $i,j=1,2,3.$) denotes the component \(C_{ij}\) of the elastic tensor matrix $C$ for microstructure \(\mathbf{s}\). \(\text{IoU}(\mathcal{A}_\mathbf{s},\mathcal{A}_\mathbf{s'})\) is the Intersection over Union operation between two microstructures ($\mathbf{s}$ and $\mathbf{s'}$), which is calculated as:
    
    \begin{equation}
        \text{IoU}(\mathcal{A}_\mathbf{s},\mathcal{A}_\mathbf{s'}) = \frac{\mathcal{A}_\mathbf{s} \cap \mathcal{A}_\mathbf{s'}}{\mathcal{A}_\mathbf{s}\cup \mathcal{A}_\mathbf{s'}}.
    \end{equation}

    Here, $\mathcal{A}_\mathbf{s} \cap \mathcal{A}_\mathbf{s'} :=\sum_{p,q} (\mathcal{A}_\mathbf{s}(p, q) \land \mathcal{A}_\mathbf{s'}(p, q))$,  $\mathcal{A}_\mathbf{s} \cup \mathcal{A}_\mathbf{s'} :=\sum_{p,q} (\mathcal{A}_\mathbf{s}(p,q) \lor \mathcal{A}_\mathbf{s'}(p, q))$ ($p,q=1,2,\cdots,256$). 
    
    After computing the similarity for each microstructure, we remove those with a similarity score greater than 0.9 from the dataset, thereby obtaining dataset $\mathcal{D}_0$ with 36,585 microstructures.
    
\end{enumerate}

\begin{figure*}[t]
    \centering
    \includegraphics[scale=0.23]{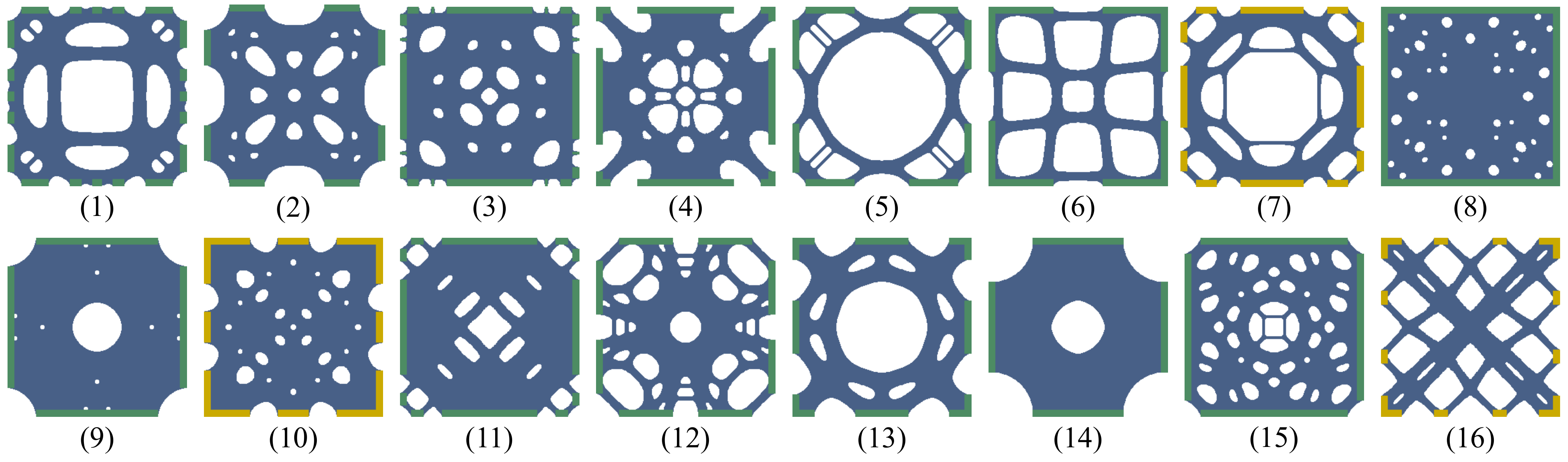}
    %16个结构
    \caption{
    %Sixteen different boundaries are obtained by the clustering method. 
    %The boundaries of all microstructures that precisely match the considered boundary are colored green, while the rest are colored yellow.
    %\tr{16 distinct boundaries resulting from clustering $\mD_0$, where green boundaries indicate the presence of corresponding microstructures within $\mD_0$, while yellow boundaries signify the absence of microstructures with those specific boundaries in $\mD_0$.}
    Illustration of sixteen boundaries derived by k-means clustering in dataset \(\mathcal{D}_0\). Green boundaries indicate that there are corresponding boundaries in \(\mathcal{D}_0\), while yellow boundaries indicate that no microstructures with these boundaries are found in \(\mathcal{D}_0\). 
    }
    \label{fig:initdata}
\end{figure*}

\paragraph{Boundary clustering}
%
%Since each microstructure is cubic symmetric, the clustering acts on the boundary of the top-left quarter of the microstructure. 
We perform clustering operations on the dataset $\mD_0$ based on microstructure boundaries.
%We perform clustering on the boundaries of microstructures of $\widehat{\mD}$. 
%Given the cubic symmetry of each microstructure, the clustering process focuses on the boundary of the top-left quarter. This region can be represented as a $128 \times 128$ matrix, and its boundary is described by a binary vector. %, as illustrated in Figure \ref{fig:similarity}. 
%A binary vector describes the boundary. 
When clustering these boundaries, we measure their similarity using the Euclidean distance. Specifically, for two boundary binary vectors $\mathbf{b}$ and $\mathbf{b}'$, the boundary similarity (BS) is calculated as:
\begin{equation} \nonumber
    \text{BS}(\mathbf{b}, \mathbf{b}') = \sqrt{\sum_{i=1}^{256} (b_{i} - b_{i}')^2}.
\end{equation}

Specifically, we use the k-means clustering method to generate a set of seed boundaries $\{\mB^{\text{seed}}_i\}$. 
%Automatically determine the number of classes?(The number of clusters is determined through experimentation. 
To determine the number of classes, we first select 40 clusters, but the clustering results show several nearly identical boundaries. 
To address this, we gradually reduce the number of clusters until there are no longer highly similar boundaries in the clustering results.
The final cluster number is 16 as shown in Figure~\ref{fig:initdata}. Notably, the k-means clustering process is highly efficient, with all operations in this study completed in two minutes.

\begin{figure}[t]
    \centering
    \includegraphics[width=0.9\linewidth]{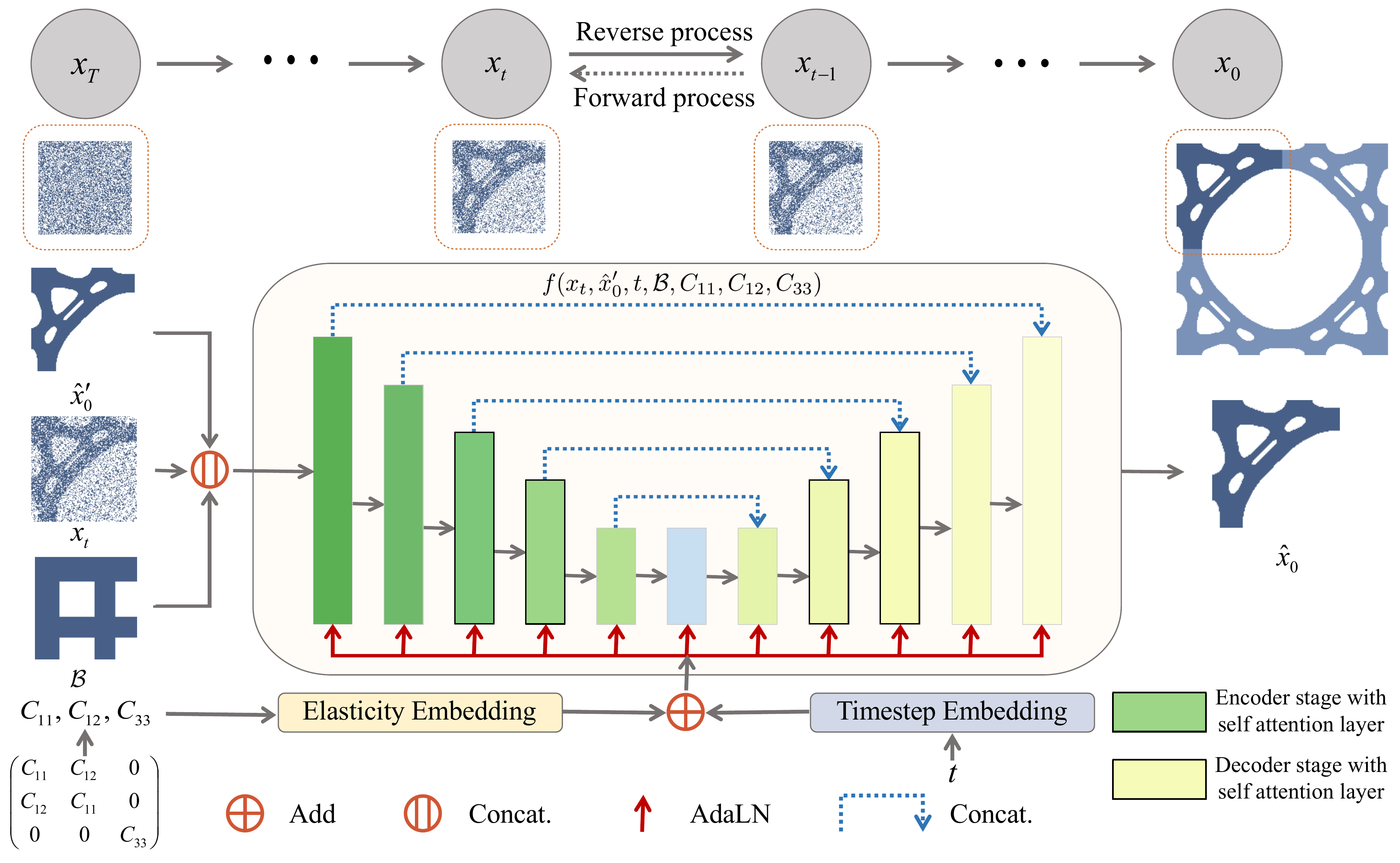}
    \caption{Network architecture of the proposed self-conditioning diffusion model. %\tr{change the 1-st and 5-th microstructure}
    }
    \label{fig:network}
\end{figure}

\paragraph{Boundary-constrained generation}
Using the dataset $\mD_0$, we train a self-conditioning diffusion model that is capable of generating microstructures given specified elastic tensors and boundary types, as illustrated in Figure~\ref{fig:network}. A detailed description of this model is provided in Section~\ref{sec:network}. 
For each seed boundary $\mB^{\text{seed}}_i$, we run the generative model to generate various microstructures to form a dataset $\hat{\mD}$ satisfying the boundary-identical and cubic symmetric constraints. 
Based on the generated results, the dataset $\hat{\mD}$ has limitations regarding its elastic moduli space. 
To this end, we propose an iterative algorithm, an active learning strategy, to alternate model retraining and dataset augmentation.
%\tb{the overall network illustration...}

\subsection{Self-conditioning diffusion model with property guidance} \label{sec:network}
%\tr{Since the microstructures are cubic symmetric, the top-left quarter of the microstructure $x$ is utilized during the training and generation. It is represented by the first 128 rows and 128 columns of $\mathcal{A}$. Besides, we use symmetric Gaussian noise to ensure that the generated microstructures adhere to cubic symmetry.} 
Given the cubic symmetry of the microstructure, we utilize the top-left quarter as a representative proxy for the entire structure. It facilitates computational resources during network training and microstructure generation. This portion is designated as \( x \) and is represented by the first 128 rows and 128 columns of matrix \( \mathcal{A} \). Furthermore, to ensure that the generated microstructures exhibit cubic symmetry, we utilize symmetric Gaussian noise in the generation process.
\paragraph{Diffusion Model}
Diffusion models typically serve to denoise Gaussian noise $x_T$ towards a data sample $x_0$ in $T$ steps, involving both forward and reverse processes during training (Figure~\ref{fig:network}).

\begin{itemize}
    \item \textbf{Forward Process}: This process begins with a data sample $x_0$ and progressively generates a noisy sample $x_t$ at each time step $t$. The addition of noise is achieved by sampling a Gaussian noise $\epsilon \sim \mathcal{N}(0, I)$. 
    \begin{equation}
    x_t = \sqrt{\gamma_t} x_0 + \sqrt{1 - \gamma_t} \epsilon,
    \end{equation}
    $\gamma_t$ is a noise schedule that gradually changes from 1 to 0. At each step, the model adds a bit of this Gaussian noise to the data sample, thereby gradually transforming the original data $x_0$ into a noisy version $x_t$ over $T$ steps.
    \item \textbf{Reverse Process}: The reverse process commences from pure Gaussian noise $x_T \sim \mathcal{N}(0, 1)$. The model is trained to learn the denoising sequence, effectively reversing the noising process step by step until it reaches $x_0$. In each step of this process, the model takes a noisy sample $x_t$ and estimates the less noisy sample $x_{t-1}$. %We generally use the neural network to predict this estimation, where the denoising process takes $x_t$ and estimates $x_{t-1}$ by inferring $x_{t-1}$, $\epsilon$, or $x_0$~\cite{ho2020denoising}.
    We generally use a neural network to predict this estimation. However, instead of directly learning the transition from $x_t$ to $x_{t-1}$, we employ a network, denoted as $f(x_t,t)$, to predict $x_0$ from $x_t$. We then estimate $x_{t-1}$ based on $x_t$ and the predicted $\hat{x}_0$. Following~\cite{ho2020denoising}, the backbone of this denoising network is implemented as a U-Net architecture shown in Figure~\ref{fig:network}.
\end{itemize}

\paragraph{Self-conditioning}
Self-conditioning is a technique proposed by Chen et al.\cite{chen2022analog} to improve the generation quality of diffusion models. The core concept of self-conditioning involves utilizing the model's previously generated partial outputs as supplementary information to guide subsequent generation steps. This approach enhances the model's capacity to leverage conditional information, resulting in data that aligns more closely with the conditional input. The specific approach is as follows: we consider a slightly different denoising function, $f(x_t,\hat{x}'_0,t)$, which takes the concatenation of the previous estimation $\hat{x}'_0$ for $x_0$ and $x_t$ as input to the network.  Building on the work of Chen et al.~\cite{chen2022analog}, during the training process, we estimate $\hat{x}'_0 = f(x_t, 0, t)$ with a 50\% probability and use this estimate for self-conditioning. For the remaining instances, we set $\hat{x}'_0$ to 0, effectively reverting to a modeling approach without self-conditioning. %As suggested by Chen et al.\cite{chen2022analog}, during the training process, % we set $\hat{x}'_0$ to $f(x_t, 0, t)$ with probability 50\%, and at other times, $\hat{x}'_0$ is set to 0, i.e., without self-conditioning.
%\tr{with a probability of 50\%, we first estimate $\hat{x}'_0 = f(x_t, 0, t)$ and then use it for self-conditioning, at other times, we set $\hat{x}'_0$ to 0 which falls back to modeling without self-conditioning.}

\paragraph{Conditions}%Representations and conditions}
%We represent each microstructure using a binary matrix of $256\times256$, where 0 represents empty space and 1 represents solid space. \tb{
%The reason we chose a resolution of $256\times256$ is that a matrix at this resolution is sufficient to represent microstructures with a rich variety of physical properties and geometric shapes. Moreover, the computational efficiency at this resolution is sufficiently high.
%The resolution of $256\times256$ effectively represents diverse microstructures and offers satisfactory computational efficiency. 
%} 
%具体实现如下：我们首先对$C_{11}$， $C_{12}$和$C_{33}$分别进行归一化，然后使用可学习的正弦嵌入对归一化后的三个量进行编码。这里使用的可学习的正弦嵌入是一种使用正弦嵌入~\cite{attention}作为初始化的可学习的位置编码，其具体公式如下：\[ PE_{(pos, 2i)} = \sin\left(pos \cdot w_i \cdot 2\pi\right) \]
%\[ PE_{(pos, 2i+1)} = \cos\left(pos \cdot w_i \cdot 2\pi\right) \]

%where PE stands for Positional Embedding, \( w_i \) is a learnable parameter initialized as:\[ w_i = \frac{1}{10000^{2i/d}} \]。
%\( d \) represents the dimensionality of the embedding vector。
%然后，我们将三个编码向量与时间步嵌入相加，并通过自适应层归一化将结果合并到 U-Net 的每一层~\cite{adaLN}。根据~\cite{ho2021classifier}，在训练过程中，我们以20\%的概率随机将弹性张量条件设置为空，并在生成过程中将引导尺度设置为1。

Our self-conditioning diffusion model aims to generate microstructures based on the given boundary and elastic tensor. For microstructures with cubic symmetry, their homogenized elastic tensor $C^H =[C_{11},C_{12},C_{13}; C_{21},C_{22},C_{23}; C_{31},C_{32},C_{33} ]$ can be fully determined by three independent components: $C_{11}$, $C_{12}$, and $C_{33}$. Therefore, we only consider these three components as the conditions for elastic tensor. We incorporate them into the network using classifier-free guidance~\cite{ho2021classifier}. %Here is the specific implementation: We encode $C_{11}$, $C_{12}$, and $C_{33}$ using learnable sinusoidal embeddings~\cite{attention}. 
The implementation is detailed as follows: Initially, we normalize $C_{11}$, $C_{12}$, and $C_{33}$. Following this normalization, the three quantities are encoded using Learnable Sinusoidal Embedding (LSE). The LSE utilized here is a type of learnable positional encoding that initializes with Sinusoidal Positional Embedding (SPE)~\cite{attention}. The specific formula for this embedding is defined as:
   \begin{align}
        PE_{(pos, 2i)} &= \sin\left(pos \cdot w_i \cdot 2\pi\right), \\
        PE_{(pos, 2i+1)} &= \cos\left(pos \cdot w_i \cdot 2\pi\right).
    \end{align}
Here, $PE$ denotes Positional Embedding, and $w_i$ represents a learnable parameter initialized as: $ w_i = \frac{1}{10000^{2i/d}}$, 
$d$ stands for the dimensionality of the embedding vector.

Subsequently, the three encoded vectors are added to the timestep embedding, and the combined result is integrated into each layer of the U-Net through Adaptive Layer Normalization (AdaLN) \cite{adaLN}. According to \cite{ho2021classifier}, during training, we randomly set the elastic tensor condition to null with a probability of 20\%, and during the generation process, the guidance scale is set to 1.

%Then, we add the three encoded vectors with the timestep embedding and incorporate the result into every layer of the U-Net through adaptive layer normalization~\cite{adaLN}. According to~\cite{ho2021classifier}, during training, we randomly set the elastic tensor condition to empty with a probability of 20\% and set the guidance scale to 1 during generation.

For the boundary constraint, we encode the boundary information into an image and then append this image to the network's input. Specifically, assuming that we want to use the boundary $\mB$ of a particular microstructure as a condition, we first extract %a \(128 \times 128\) sub-microstructure image from the top-left quarter of the given microstructure
its top-left quarter and record the positions of the rows and columns where the upper and left boundaries lie. 
Then, a new binary image is created (default value 0) where the recorded boundary rows and columns are set to 1, establishing a boundary representation for the given microstructure.

\paragraph{Network architecture}
The Encoder and Decoder parts of our U-Net both consist of 5 stages. We set the number of resblocks for each stage as 1, and channel multiplications for each stage are set to 1, 2, 4, 8, and 8, respectively. The input and output feature dimensions for each layer are as follows: (32,32), (32,64), (64,128), (128,256), (256,256), (256,256), (256,256), (256,128), (128,64), (64,32), (32,32). Attention resolutions with respect to feature map sizes are set at 4 and 8. The diffusion step number is set to 1,000 with a linear noise schedule during training.

\paragraph{Training details}
The network training is based on the following denoising
loss:
\begin{equation}
    \mathcal{L}_{x_0}=E_{\epsilon\sim\mathcal{N}(0,I),t\sim \mathcal{U}(0,I)}||f(x_t,\hat{x}'_0,t,\mB,C_{11},C_{12},C_{33})-x_0||_2^2.
\end{equation}
We use the AdamW optimizer~\cite{kingma2014adam,loshchilov2017decoupled} to train the diffusion model. 
For each model mentioned below, we fix the learning rate to $10^{-4}$, batch size to 256, and train for 1000 epochs.
%We show the loss function as a function of the iteration number in Figure~\ref{fig:training}.
\begin{figure*}[!b]
    \centering
    \includegraphics[scale=0.23]{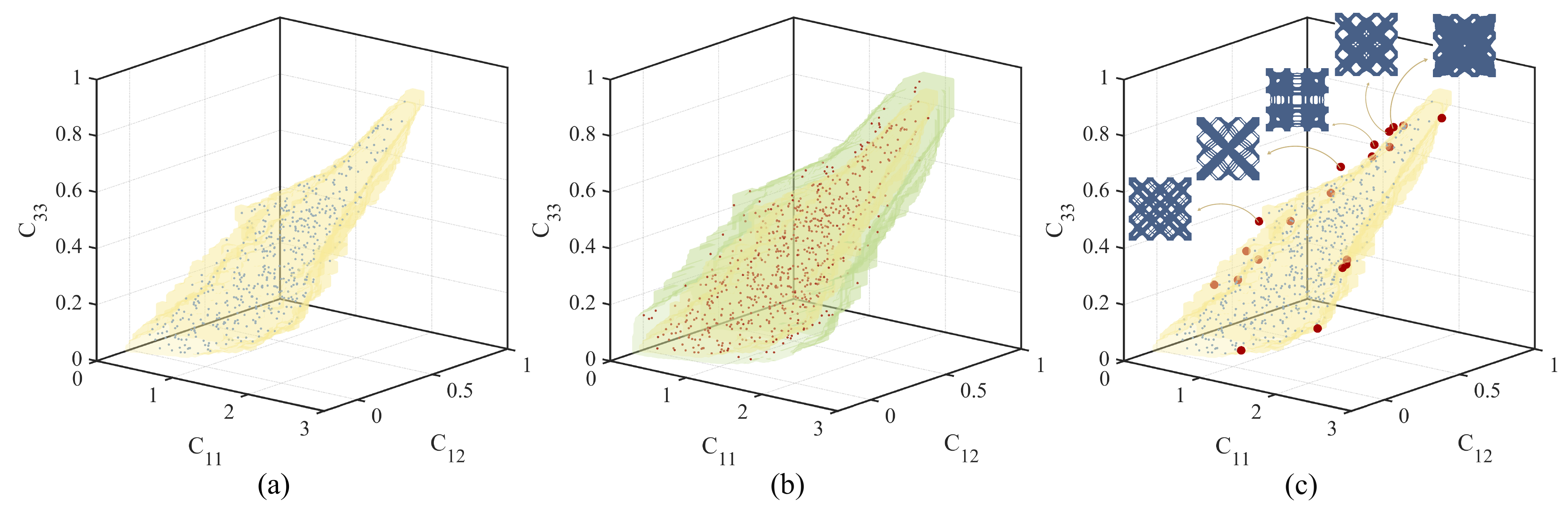}
    \caption{Illustration of the generation process: (a) Interior of the Signed Distance Function (SDF) representing the training set's property coverage. (b) The expanded SDF (green region) and the target elastic tensor (red points). (c) Microstructures generated by the network that lie beyond the coverage of the training set properties.}
    \label{fig:generation}
\end{figure*}

\begin{figure*}[t!]
  \centering
  \begin{overpic}[width=1\linewidth]{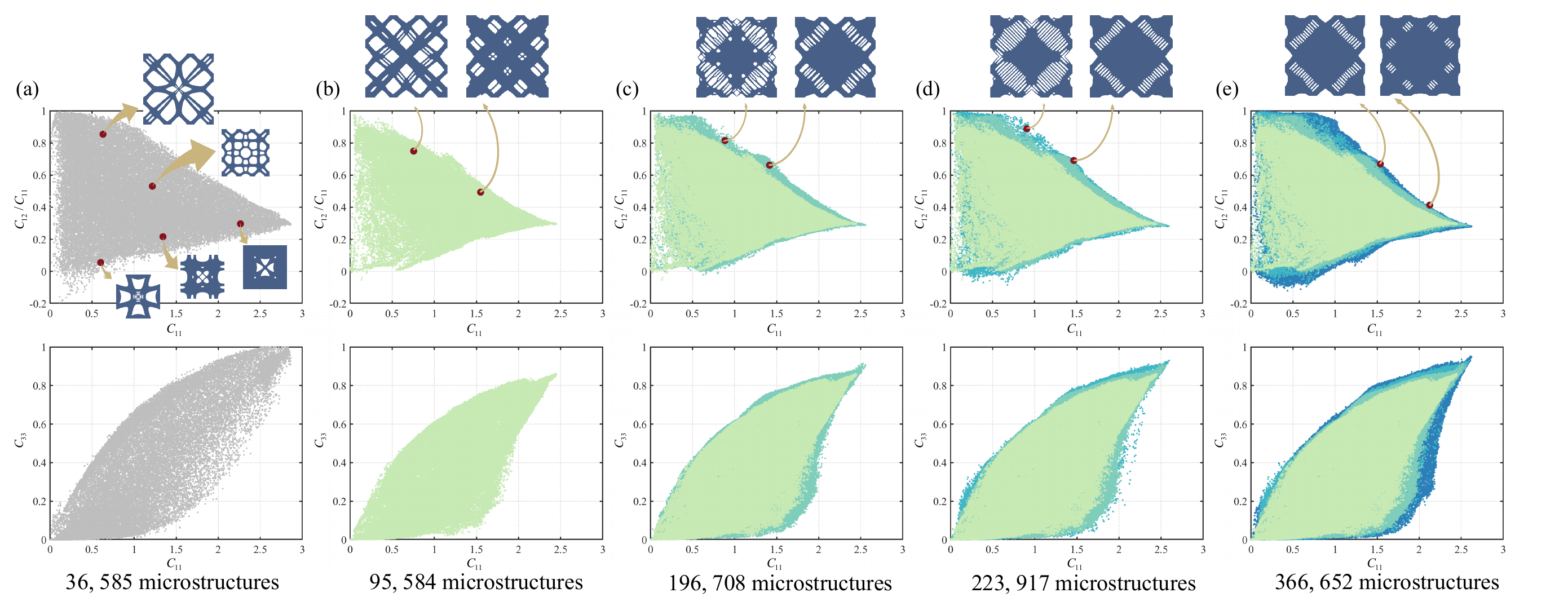}
    {
      % \put(72,-3){\small Ours}
      % \put(19,0){\scriptsize $N = 24$}
    }
  \end{overpic}
  \vspace{0mm}
  \caption{
  Overview of dataset generation framework: 
  (a) Dataset $\mathcal{D}_0$ with microstructures exhibit diverse boundaries. 
  (b) One identical boundary dataset constructed from the diffusion model with a single seed boundary.
  (c)-(d) Active learning process. (e) The final expanded dataset. 
  }
  \label{fig:pipeline}
\end{figure*}

\paragraph{Generation}
We use the diffusion model to generate diverse microstructures under two conditions: (1) the specified boundary and (2) the predefined elastic tensor matrix.
To ensure the diversity of microstructures generated by the network, we utilize a set of elastic tensors with a broad range of values as conditions during sampling. To create this set of elastic tensors that encompasses a wide and reasonable range, we first compute the SDF values at each node of a Cartesian grid covering the property space of the training dataset \cite{green2007improved,mishchenko2015fast}, as illustrated in Figure \ref{fig:generation} (a).  Then, we set the SDF values to negative for the small neighborhood of nodes where the SDF is less than 0. This ensures that the sampling range is not limited to the coverage range of the training dataset. Finally, we perform sampling in the regions where the SDF is less than 0 to obtain the elastic tensors required for generation. Figure ~\ref{fig:generation}(b) displays the expanded range of the SDF and the elastic tensors obtained from the sampling. We input these elastic tensors and the encoding of the target boundary together as conditions into the network and use 50 steps of DDPM sampling for microstructure generation. %(shown in Figure~\ref{fig:gener})
 Figure ~\ref{fig:generation}(c) shows microstructures generated by our generative model that lie outside the %coverage range of propertys in the training set.
space of the training set. 

\paragraph{Active learning strategy}
Building upon the aforementioned generation process, we can iteratively expand a boundary-identical dataset, as depicted in Figure ~\ref{fig:pipeline}. This expansion involves incorporating the generated microstructures into the existing dataset to form an augmented training set, which is used to further train the network and produce subsequent generations. Through this iterative augmentation process, we can construct a large-range boundary-identical microstructure dataset.
%\tr{The whole dataset generation process is shown in Figure~\ref{fig:pipeline}. 
%Thus, a large-range boundary-identical microstructure dataset is built. }

\subsection{Multiscale topology optimization}

When a dataset with large-range identical boundaries, denoted as $\mathcal{D}$, is constructed, multiscale topology optimization becomes significantly more efficient, as it can be implemented directly without concerns regarding connectivity issues. We illustrate the multiscale optimization model using the invisibility cloak ~\cite{wang2022mechanical} as an example. 
The reference design space is denoted as $\Omega$, and the mechanical cloak is defined at the inside of $\Omega$ and marked $\Omega_c$. The properties of the cloak are evaluated in the remaining areas $\Omega_s = \Omega / \Omega_c$, and the stealth capabilities of the cloak are deemed better when the difference between the displacement and the reference displacement is minimized. 
The elastic tensors, $\mathbf{C}$, are employed as the design variables. The objective function is defined as the deviation between the calculated displacement $\mathbf{u}_o$ and the targeted displacement $\mathbf{u}_t$ in $\Omega_s$. The constraints are determined by the projection of the equilibrium equation and the elastic tensor within the design space, $\mathcal{S}$. The optimization model is formulated as follows:
%; see~\cite{wang2020deep} for details. 
\begin{align}
    \min_{\mathbf{C}}\, &\| \mathbf{u}_o - \mathbf{u}_t \|_2^2 \\
    \text{s.t. } \,&\mathbf{K}(\mathbf{C})\mathbf{u} = \mathbf{F}, \\
      & \varphi(\mathbf{C}_e,f_\mathcal{S}) \leq 0. \quad \forall e;
\end{align}    
Here, $\mathbf{K}(\mathbf{C})$ represents the stiffness matrix, and $\mathbf{F}$ denotes the external force vector. The function $\varphi(\cdot, \cdot)$ indicates that the elastic tensor $\mathbf{C}_e$ of macrostructure element $e$ ensures the SDF ($f_\mathcal{S}$) remains less than zero. %where $\mathbf{K}(\mathbf{C})$ is the stiffness matrix. %$\mathbf{v}$ represents a binary vector indicating nodes of interest for displacement in the finite element mesh; $\mathbf{u}_t$ is the target displacement field; 
%$\mathbf{f}$ is load vector. $\varphi(\cdot,\cdot)$ represents the elastic tensor C causes the symbolic distance function constructed by S to be less than 0.} %is the constraint function for the macroscopic elastic tensor distribution within the ~\cite{wang2020deep}.}
%Next, let $\mathbf{C}_e$ be the optimized elasticity tensor corresponding to element $e$.  
Then, we find $\mathbf{s}_e \in \mathcal{D}$ that minimizes: 
\begin{equation}
   \mathbf{s}_e = \arg\min_{\mathbf{s} \in \mathcal{D}} \| \mathbf{C}(\mathbf{s}) - \mathbf{C}_e \|_2,
\end{equation}
Finally, optimized microstructure $\mathbf{s}_e$ is obtained and assembled into the final multiscale structure.

\section{Numerical experiments and discussion}\label{sec:res}

We validate the proposed microstructure datasets through various numerical examples. %We start by discussing boundary-identical properties in Section~\ref{sec:bound}. Then network performance, including the accuracy of the generated structure, is analyzed in Section~\ref{sec:npa}. 
%The coverage of the moduli of the sixteen boundary-identical microstructure datasets are shown in Section~\ref{sec:bound-ana}. Examples of multiscale optimization design and their efficiency statistics are analyzed in Section~\ref{sec:multiscale}. 
%All our experiments are done on a server with 2 CPUs (Intel Xeon Silver 4316 2.30GHz), 512GB RAM and 8 Nvidia GeForce RTX 3090 GPUs. 
The training and inference of our network, as well as the full-scale finite element analysis for multi-scale structures, are conducted on a server equipped with 2 CPUs (Intel Xeon Silver 4316 2.30GHz), 512GB of RAM, and 8 Nvidia GeForce RTX 3090 GPUs. The multi-scale design process is carried out on a computer with 1 CPU (Intel i7-6700K 4.00GHz) and 16GB of RAM. The material has Young's modulus of $E = 3$ and Poisson's ratio of $\nu = 0.3$. Linear elements are used to perform finite element analysis.

\subsection{Identical boundaries discussion}
\label{sec:bound}

\paragraph{Compound microstructures testing} 
%The boundary connectivity between microstructures plays a critical role in determining the mechanical properties of the resulting composite structures. 
%To evaluate the mechanical compatibility of the boundary-identical microstructures we designed, we compared two sets: one with boundary-identical microstructures and another with microstructures that are only partially connected at the boundaries. 
%We then plotted the bulk modulus for both individual cells and compound states alongside the theoretical Hashin-Shtrikman (HS) bounds. 
%As shown in Figure~\ref{fig:comp1}, while all the selected single cells approach the HS bounds, the compound structures formed by microstructures with partial boundary connections exhibit a significant gap between their bulk modulus and the HS bounds. 
%In contrast, compound structures formed by boundary-identical microstructures closely align with the HS bounds. This behavior is due to the extent of boundary connectivity, which directly influences mechanical performance. When the boundaries are only partially connected, overall performance is degraded.

The boundary connectivity between microstructures plays a critical role in determining the mechanical properties of the resulting composite structures. To evaluate the mechanical compatibility of the boundary-identical microstructures we design, we compare two sets: one with boundary-identical microstructures and another with microstructures that are only partially connected at the boundaries.

\begin{figure*}[t]
    \centering
    \includegraphics[scale=0.23]{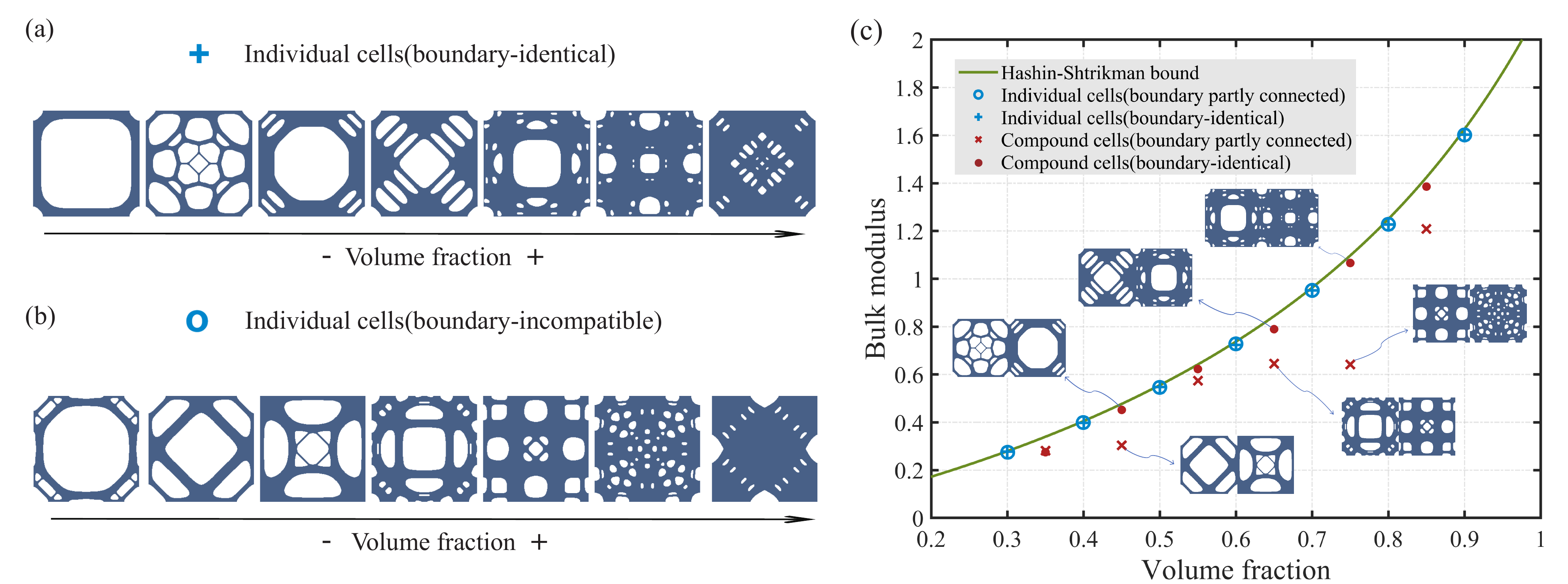}
    \caption{(a) A set of individual cells with identical boundaries that achieve the Hashin-Shtrikman (HS) upper bound, with volume progressively increasing from left to right. (b) A set of individual cells that also reach the HS upper bound but have partially connected boundaries, similarly showing a progressive increase in volume from left to right. (c) The relationship between the bulk modulus and volume fraction for both sets of individual cells, as well as for compound cells formed by adjacent cells within the same group. }%\tb{The relationship between the bulk modulus and volume fraction for microstructures with different boundary connections and their compound formulation. \tr{show the tested microstructure set}}}
    \label{fig:comp1}
\end{figure*}

We plot the bulk modulus for both individual cells and compound states alongside the theoretical Hashin-Shtrikman (HS) bounds. As illustrated in Figure~\ref{fig:comp1}, while all selected single cells approach the HS bounds, the compound structures formed from microstructures with partial boundary connections exhibit a significant gap between their bulk modulus and the HS bounds. In contrast, compound structures composed of boundary-identical microstructures closely align with the HS bounds. This behavior arises from the degree of boundary connectivity, which directly influences mechanical performance. When boundaries are only partially connected, overall performance degrades.

%The boundary connectivity between microstructures significantly influences the mechanical properties of the resulting compound structures. To verify that the boundary-identical microstructures we constructed possess good mechanical compatibility, we selected a set of boundary-identical microstructures and a set of microstructures which boundaries partly connected, and plotted their bulk modulus in both single-cell and compound states, along with the theoretical Hashin-Shtrikman (HS) bounds. From Figure ~\ref{fig:comp1}, it can be observed that although the selected single cells are all capable of approaching the HS bounds, for the compound outcomes formed by single cells which boundaries partly connected, there remains a considerable gap between their bulk modulus and the HS bounds. In contrast, compound structures formed by boundary-identical microstructures maintain their proximity to the HS bounds. This is propertyd to the fact that connection at the boundaries affects the mechanical performance at the interfaces, leading to a degradation in the overall performance of the compound structures.}

\begin{figure*}[!t]
    \centering
    \includegraphics[scale=0.38]{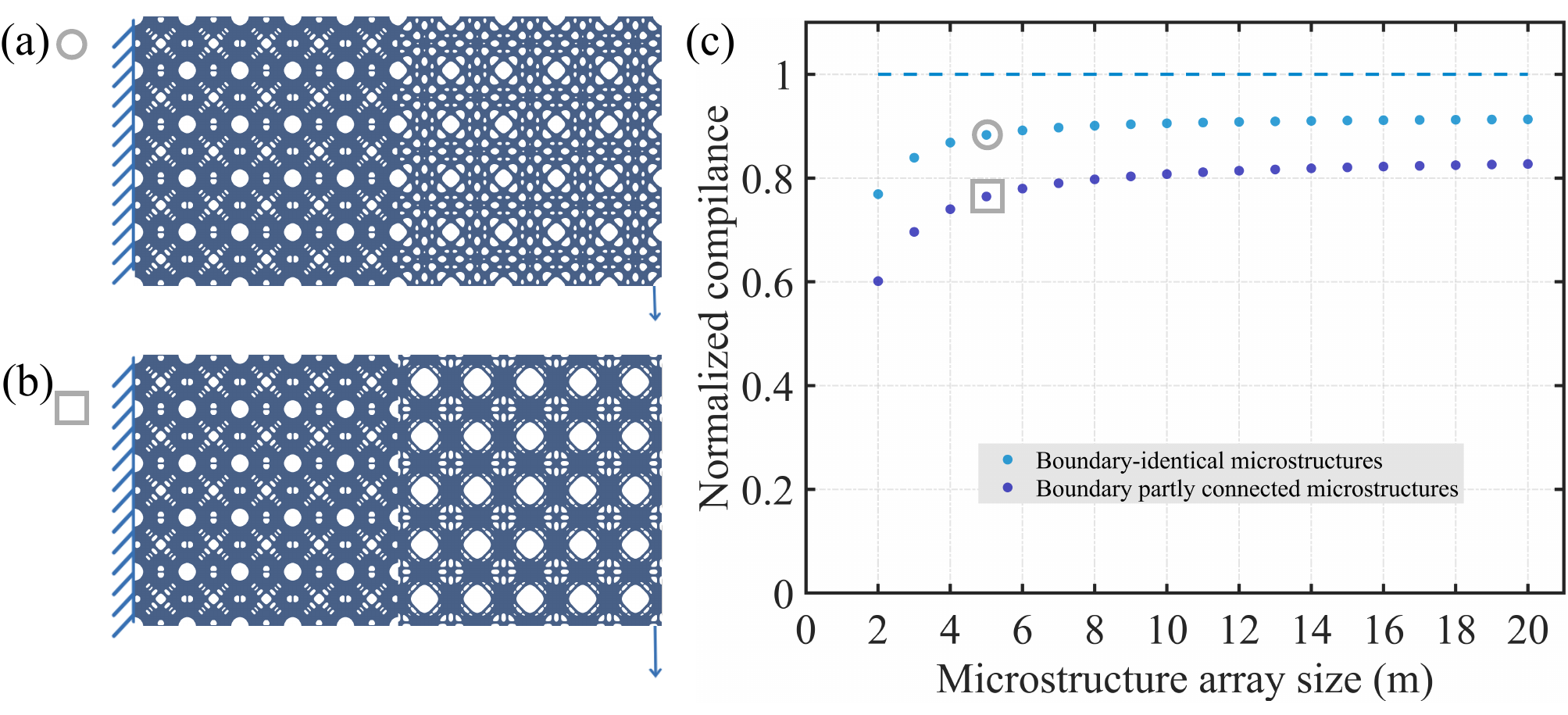}
    \caption{A cantilever beam with two microstructural regions: (a) filled with boundary-identical structures (b) filled with partly connected boundary structures. (c)The relationship between normalized compliance and the array size $m$. }
    \label{fig:comp2_2}
\end{figure*}

\begin{figure*}[!t]
    \centering
    \includegraphics[scale=0.26]{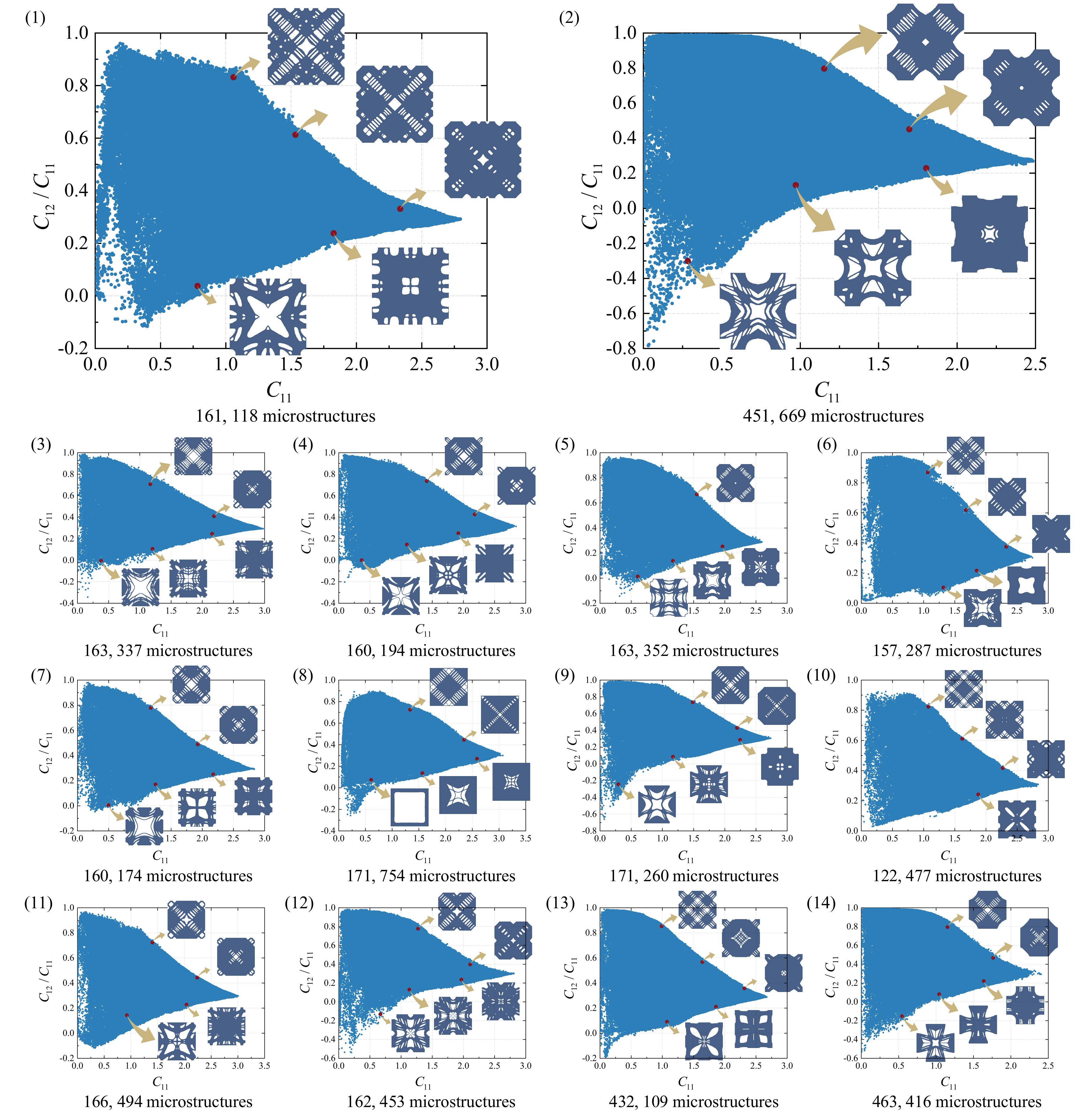}
    \caption{The generated 14 large-range boundary-identical microstructure datasets.}
    \label{fig:datasets}
\end{figure*}
\paragraph{Full scale analysis} 
%The multiscale design, based on homogenization theory, assumes scale separation, meaning that the theory evaluates the properties of an infinitely periodic array of microstructures. 
%However, achieving such an infinite array in practical manufacturing is impossible, leading to inevitable discrepancies between the properties predicted by homogenization theory and those of actual multiscale structures. 
%To examine the impact of boundary connectivity between structures, we conducted an experiment involving a cantilever beam composed of two regions, each filled with \( m \times m \) identical microstructures, as illustrated in Figure~\ref{fig:comp2_2}. 
%In Figure~\ref{fig:comp2_2}(b), the microstructures are only partially connected at the boundaries, whereas in Figure~\ref{fig:comp2_2}(a), the boundaries are identical. In both cases, the repeating number is 5. 
%Importantly, the structures filling the same regions in both cantilever beams possess identical elastic tensors. Figure~\ref{fig:comp2_2}(c) plots the normalized compliance against $m$, where normalized compliance is defined as the ratio of compliance from homogenization analysis to that from full-scale analysis. 
%As shown in Figure~\ref{fig:comp2_2}(c), the normalized compliance approaches 1 as the resolution parameter $m$ increases. 
%Moreover, structures with identical boundaries exhibit higher normalized compliance, indicating that multiscale designs utilizing more boundary-identical microstructures yield properties closer to those predicted by homogenization theory.

The multiscale design, based on homogenization theory, assumes scale separation, meaning that the theory evaluates the properties of an infinitely periodic array of microstructures. However, achieving such an infinite array in practical manufacturing is not possible, leading to inevitable discrepancies between the properties predicted by homogenization theory and those of actual multiscale structures.

To examine the impact of boundary connectivity between structures, we conduct an experiment involving a cantilever beam composed of two regions, each filled with \( m \times m \) identical microstructures, as illustrated in Figure~\ref{fig:comp2_2}. In Figure~\ref{fig:comp2_2}(b), the microstructures are only partially connected at the boundaries, while in Figure~\ref{fig:comp2_2}(a), the boundaries are identical. In both cases, the repeating number is 5. Importantly, the structures filling the same regions in both cantilever beams possess identical elastic tensors.

Figure~\ref{fig:comp2_2}(c) plots the normalized compliance against \(m\), where normalized compliance is defined as the ratio of compliance from homogenization analysis to that from full-scale analysis. As shown in Figure~\ref{fig:comp2_2}(c), the normalized compliance approaches 1 as the resolution parameter \(m\) increases. Structures with identical boundaries exhibit higher normalized compliance, indicating that multiscale designs utilizing more boundary-identical microstructures yield properties closer to those predicted by homogenization theory.

\paragraph{Boundary-identical microstructure datasets analysis} 
\label{sec:bound-ana}

We present the property coverage of 16 boundary-identical large-range microstructure datasets shown in  Figures~\ref{fig:datasets} (datasets 1-14), \ref{fig:one-result} (dataset 15), and \ref{fig:pipeline} (dataset 16).  Each dataset contains over 100,000 microstructures. These datasets can be used for multiscale design or other data-driven researches. 
We conduct an analysis of the bulk modulus of all microstructures in the aforementioned datasets and compare them with the Hashin-Shtrikman bound (HS-bound)~\cite{hashin1963variational}. The structures achieve 98\% or higher of the HS-bound. Details are shown in Figure~\ref{fig:hsbound}.  This indicates that our model has the potential to generate extreme materials with specified boundaries.

\begin{figure*}[t]
    \centering
    \includegraphics[scale=0.36]{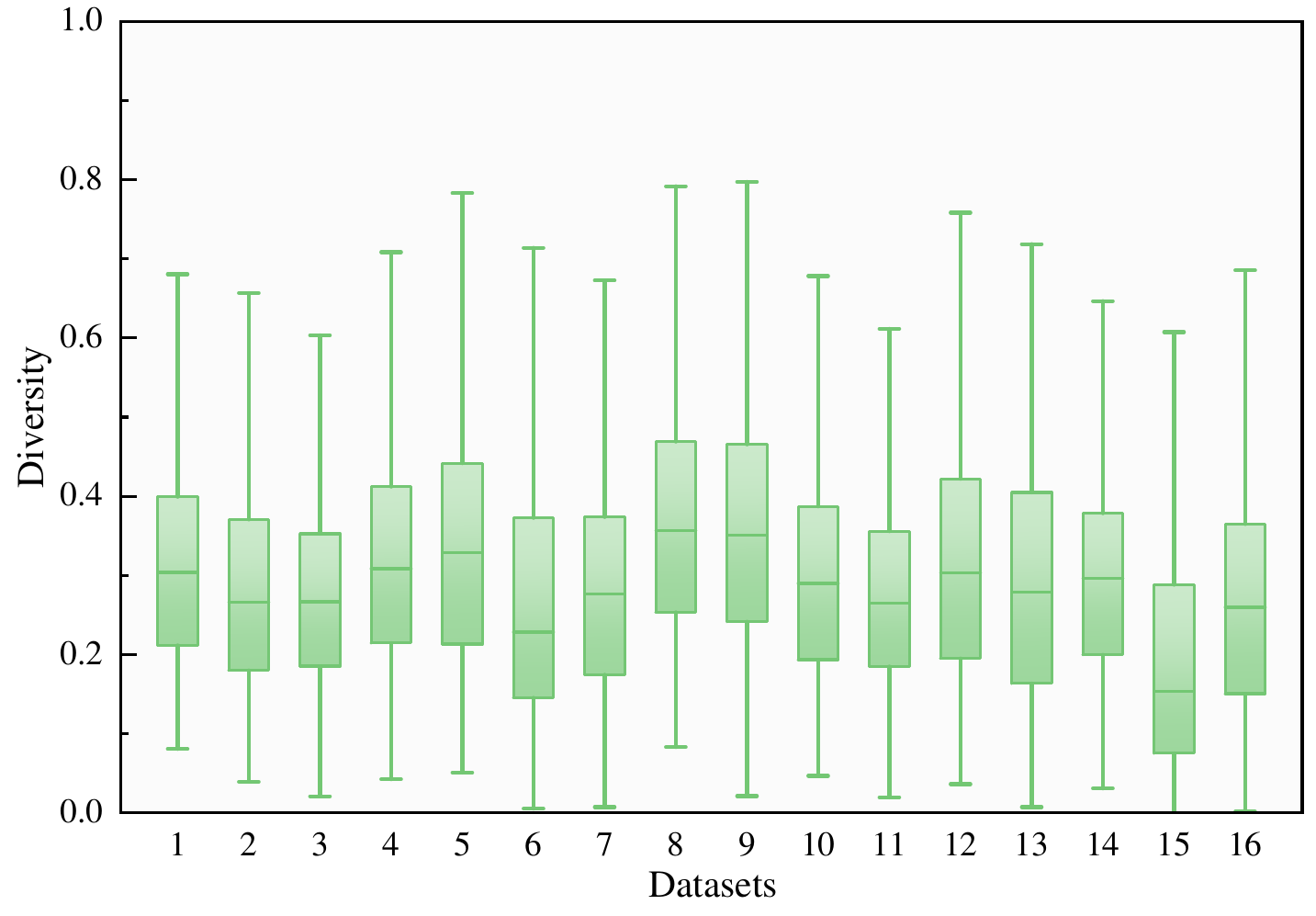}
    \caption{The diversity statistics across the 16 datasets.}
    \label{fig:fig_box}
\end{figure*}

To further analyze the diversity of microstructures, we focus on the novelty of the generated microstructures relative to those in the dataset $\mathcal{D}_0$. The diversity here is primarily assessed by comparing the novel data generated by the network with the microstructures contained in the dataset $\mathcal{D}_0$.
Thus, the structure diversity (SD) is defined as:
\begin{equation}
\text{SD}(\mathbf{s}) = 1-\text{SS}(\mathbf{s}). %= 1-\max \left( \{ \text{IoU}(\mathcal{A}_\mathbf{s},\mathcal{A}_\mathbf{s'}) : \mathbf{s'} \in N(\mathbf{s}) \}  \right). 
\end{equation}
where SS($\mathbf{s}$) is the structure similarity defined in Eq.~\eqref{eq:ss}. The value signifies that as it approaches 0, the current dataset becomes more similar to the reference dataset $\mathcal{D}_0$, whereas a value closer to 1 indicates greater novelty. Figure~\ref{fig:fig_box} presents the diversity results across 16 datasets.

\subsection{Network performance analysis}
\label{sec:npa}

To assess the capability of the self-conditioning diffusion model to generate microstructures under specified conditions, we perform conditional generation tests on a network that has been trained using the dataset $\mathcal{D}_0$. This dataset comprises various microstructures characterized by diverse boundaries $\mathcal{B}$ and elastic tensors $C$. The diversity within the training data enhances the model's generalization ability and presents a significant challenge in generating microstructures that strictly conform to specified boundaries and elastic tensors. Consequently, conducting conditional generation tests on this trained network can effectively reveal the true performance of the model structure.

We evaluate the network's performance under all 16 boundaries, using two primary metrics: (1) the $R^2$-score between the generated elastic tensors and the target elastic tensors, which is defined as:
\begin{equation}
   R^2(\mathbf{Y},\mathbf{Y}') = 1 - \frac{\sum_{i=1}^{N} ||\mathbf{y}_i - \mathbf{y}'_i||^2}{\sum_{i=1}^{N} ||\mathbf{y}_i - \bar{\mathbf{y}}||^2},
\end{equation}
where $\mathbf{Y}$ %( =\{y_1,\cdots,y_N\})$ 
and $\mathbf{Y}'$ % (=\{y_1',\cdots,y_N'\})$ 
are the generated %(e.g., $C_{11}$, $C_{12}$, $C_{33}$) 
and target elastic tensor matrix, respectively, each with the dimension of $k\times N$. $N(=1000)$ is the number of generated microstructures. $\mathbf{y}_i$ is the $i$-th column of $\mathbf{Y}$. %$\mathbf{y}'_i$ is the $i$-th column of $\mathbf{Y}'$.
For $k=1$, $\mathbf{y}_i$ corresponds to a single elastic tensor component ($C_{11}$ or $C_{12}$ or $C_{33}$) of the $i$-th generated microstructure.
When $k=3$, $\mathbf{y}_i$ is the elastic tensor vector ([$C_{11}$, $C_{12}$, $C_{33}$]) of the $i$-th generated microstructure. $\mathbf{y}'$ follows a similar explanation. 
%\tb{$\mathbf{y}'$ has the similarly meaning for $k=1$ and $k=3$. }
%$\mathbf{y}=\{\mathbf{y}^{[1]},\mathbf{y}^{[2]},\cdots,\mathbf{y}^{[N]}\}$. 
%$y_i$ and $y'_i$ represent the generated property and the target property of the $i$-th microstructure. 
%the target property (e.g., the specified elastic tensor), $\mathbf{y}'_i$ denotes the actual property of the generated microstructure, 
$\bar{\mathbf{y}}$ is the mean of the generated properties. %, and $N=1000$ is the number of generated microstructures. %, with $N=1000$ in these experiments.

(2) The degree of match (DoM) between the generated boundaries and the specified boundaries, which is calculated as:
\[ \text{DoM}({\mathbf{b}},\mathbf{b}') = 1 - \frac{||\mathbf{b} - \mathbf{b}'||_2}{||\mathbf{b}||_2}, \]
where $\mathbf{b}$ represents the target boundary, and $\mathbf{b}'$ is the boundary of the generated microstructure.
The results indicate that the model performs consistently well across all 16 boundaries without exhibiting significant overfitting to any particular boundary. 
The values of $R^2$-score and DoM are shown in Tables \ref{tab:performance1}-\ref{tab:performance3}.

\begin{figure*}[!t]
    \centering
    \includegraphics[scale=0.08]{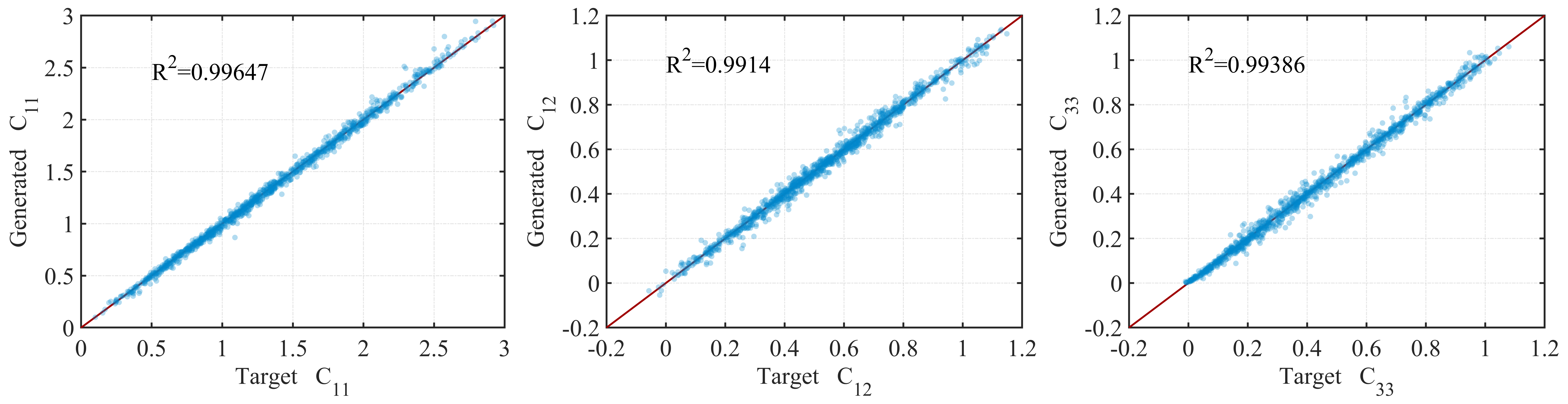}
    \caption{The $R^2$-score demonstrates the disparity between the properties of the microstructures generated by the network and the target properties.}
    \label{fig:prediction}
\end{figure*}

\begin{figure*}[!t]
    \centering
    \includegraphics[scale=0.23]{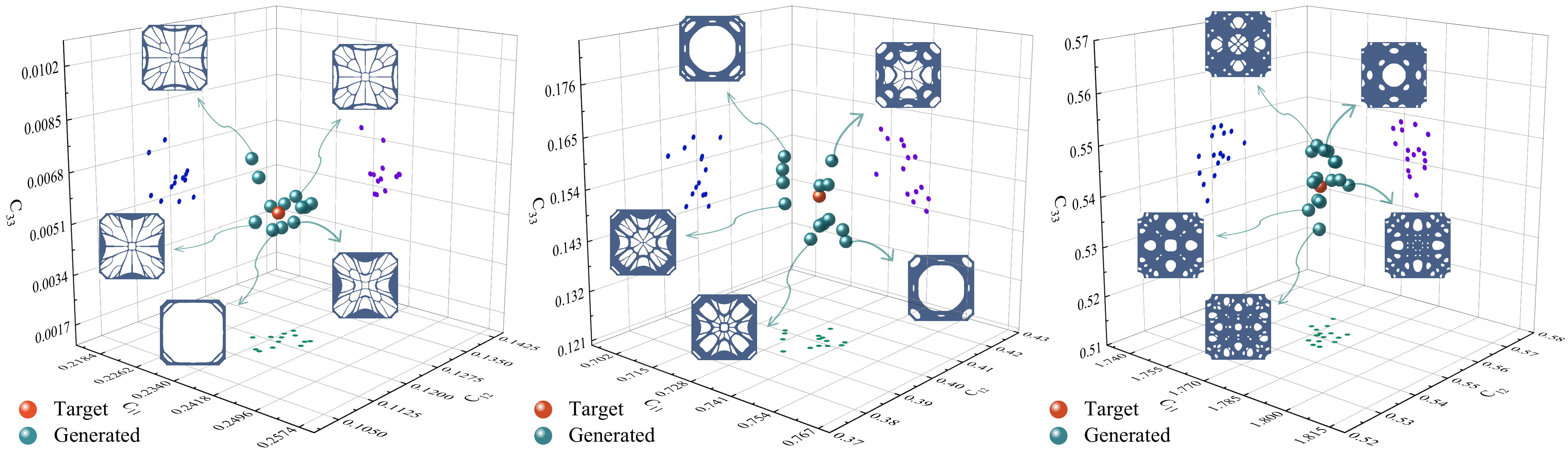}
    \caption{The generated results for three different predefined elastic tensors under the 15-th boundary.}
    \label{fig:diversity}
\end{figure*}

%and \ref{tab:performance2} present the $R^2$-scores for the test results, while Tables \ref{tab:performance3} and \ref{tab:performance4} display the degrees of match. .}

Furthermore, the test results for the 15-th dataset are visualized in Figure \ref{fig:prediction}. It shows the $R^2$-scores corresponding to three components of the elastic tensor: $C_{11}$, $C_{12}$, and $C_{33}$. These results demonstrate that the microstructures generated by the network exhibit good agreement with the target elastic tensors.

Figure~\ref{fig:diversity} illustrates the diversity of the network's generative capabilities, demonstrating that the proposed model can generate multiple structures with similar elastic tensors but distinctly different geometric configurations. By keeping the boundary conditions constant and using the same target elastic tensor across multiple generations, the results emphasize the model's ability to capture a one-to-many mapping from material properties to geometry. This highlights the model's capacity to produce a broad range of geometrically distinct microstructures that exhibit comparable elastic properties.

\subsection{Ablation study}

To investigate the impact of the self-conditioning mechanism and the encoding of elastic tensors on the network's generation performance, we design two ablation studies that focus on these aspects.

\paragraph{The ablation study on self-conditioning}

Self-conditioning has yielded excellent results in image generation tasks. To verify its effectiveness in generating microstructures, we conduct the following experiments. %\tr{For the baseline model, we adopt the architecture detailed in this article. Additionally, we train a variant of this model without the self-conditioning mechanism to serve as a control. The training dataset is selected according to the rationale provided in Section~\ref{sec:npa}, using the dataset $\mathcal{D}_0$ for training purposes. We perform generation tests under all 16 boundaries discussed in the article, evaluating the accuracy of the generated structures using the same metrics outlined in Section~\ref{sec:npa}: the $R^2$-score between the generated and target elastic tensors, and the degree of match between the generated structure's boundaries and the specified boundaries.}
For the baseline model, we use the architecture detailed in Figure~\ref{fig:network} and train a variant without the self-conditioning mechanism as a control. The training dataset, \(\mathcal{D}_0\), is selected as described in Section~\ref{sec:initdata}. Generation tests are conducted under all 16 boundaries, evaluating accuracy using the \(R^2\)-score for elastic tensors and the alignment between generated and specified boundaries.

%The $R^2$-score results are summarized in Table~\ref{tab:performance1}, whereas the matching degree results are presented in Table~\ref{tab:performance3}. 
Compared to the baseline model, the model lacking self-conditioning exhibits an average decrease of 0.0516 in $R^2$-score, with a maximum reduction of 0.1120 (Table~\ref{tab:performance1}). Moreover, there is an average decrease of 0.0255 in the DoM, with a maximum reduction of 0.7000 (Table~\ref{tab:performance3}). These test results indicate that removing the self-conditioning mechanism leads to a significant reduction in generation accuracy. This underscores the critical role of self-conditioning in enhancing the precision of the model's output.

\paragraph{The ablation study on elastic tensor embedding}

In our network, elastic tensor embeddings are integrated into each U-Net layer via AdaLN. An ablation study evaluates their impact on model performance, using the original network as the baseline. Three variants omit embeddings from the high, middle, or low layers, corresponding to the upsampling path, bottleneck, and downsampling path, respectively. All models are trained on dataset \(\mathcal{D}_0\), with sampling experiments conducted across the sixteen specified boundaries. 
%We assess the generative accuracy of the networks using the same two metrics as described in Section~\ref{sec:npa}: the $R^2$-score between the generated elastic tensors and the target elastic tensors, and the degree of match between the boundaries of the generated structures and the specified boundaries.

%
%The results for the $R^2$-score are presented in Tables~\ref{tab:performance1} and \ref{tab:performance2}, whereas the matching results are shown in Tables~\ref{tab:performance3} and \ref{tab:performance4}. 
Compared to the baseline model, the three variant models show the following reductions in $R^2$-score: an average decrease of 0.0221, 0.0398, and 0.0426, with maximum reductions of 0.8800, 0.1470, and 0.1600 (Table~\ref{tab:performance1}). Regarding the value of DoM, the average decreases are 0.0149, 0.0163, and 0.0187, with corresponding maximum reductions of 0.0380, 0.0580, and 0.0510 (Table~\ref{tab:performance3}).

Our findings suggest that omitting elastic tensor embeddings from certain layers negatively impacts the model's generative accuracy, particularly in its ability to align with the target elastic tensor. This effect is most pronounced in specific boundary sampling experiments. To maintain the accuracy and robustness of the model, it is essential to include elastic tensor embeddings in every layer of the U-Net.

\subsection{Active learning analysis}

%\paragraph{The difference between the generated elasticity tensors with original ones}

Figures~\ref{fig:generation} and~\ref{fig:pipeline} illustrate the iterative process of expanding the property space. To further quantify the computation, we first define the coverage ratio of the space and then calculate these ratios achieved through the active learning strategy.

Given that the elastic tensors we consider have only three independent components \(C_{11}\), \(C_{12}\), and \(C_{33}\), we conceptualize a three-dimensional space defined by these components, denoted as \(C_{11} \times C_{12} \times C_{33}\). The elastic tensors of all microstructures in each dataset can be visualized as a point cloud \(P\) within this 3D space. Assuming the value ranges for the three elastic tensors are \([C_{11}^{\min}, C_{11}^{\max}]\), \([C_{12}^{\min}, C_{12}^{\max}]\), and \([C_{33}^{\min}, C_{33}^{\max}]\) respectively, we divide the cuboid in this 3D space into smaller cubes using a resolution of \(N_1 \times N_2 \times N_3\), labeled as \(\{c_i\}_{i=1}^{N_1 \times N_2 \times N_3}\). We use the following metric to express the coverage of elastic tensors:
\[ \text{Range} = \frac{\# \{c_i | c_i \cap P \neq \emptyset, 1 \leq i \leq N_1 \times N_2 \times N_3\}}{N_1 \times N_2 \times N_3} .\]
This ratio indicates the proportion of small cubes that contain at least one point from the point cloud \(P\), reflecting the extent to which the elastic tensor space is covered by the dataset. In practical computations, we set the following parameters: 
 \(C_{11}^{\min} = 0\), \(C_{11}^{\max} = 3.5\),
 \(C_{12}^{\min} = -0.5\), \(C_{12}^{\max} = 1.5\),
 \(C_{33}^{\min} = 0\), \(C_{33}^{\max} = 1.5\),
 resolution: \(N_1 = 350\), \(N_2 = 200\), \(N_3 = 150\). These minimum and maximum values for the elastic tensor components are determined based on the statistical properties of all microstructures in the dataset. In the selection of resolution, we refer to the value ranges of each component to ensure that the scale intervals on the three axes are identical. Additionally, we appropriately chose the order of magnitude for the resolution to ensure that the partitioning of the elastic tensor space can effectively capture the coverage of the elastic tensors without being overly sparse or excessively dense.

% 第1到8个数据集在第0到第3次迭代过程中的属性覆盖范围
% 平均扩张6.83%，最大扩张13.87%
 On average, each iteration enhances the dataset's elastic tensor coverage by 6.83\%, with a maximum expansion of 13.87\%. Detailed statistics are provided in Table~\ref{tab:range1}. These results highlight the effectiveness of active learning methods in iteratively expanding datasets.

\begin{figure*}[b]
    \centering
    \includegraphics[scale=0.23]{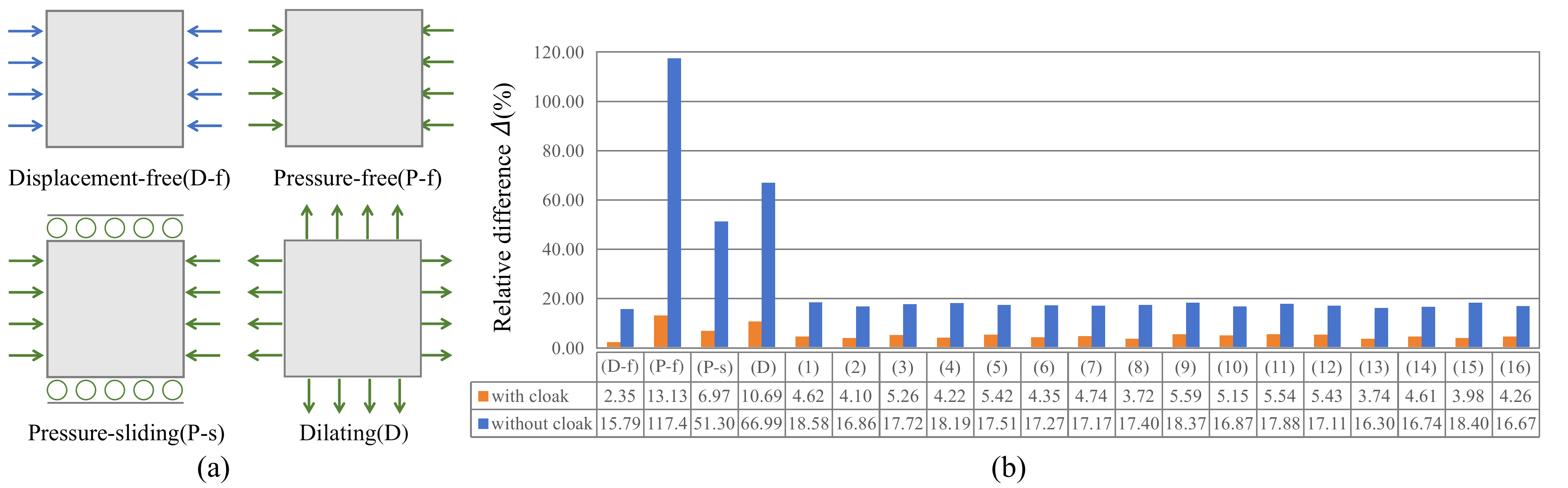}
    \caption{(a) Four initial settings. (b) Relative displacement differences of the cloak design result under different boundary conditions, void shapes, and datasets, where designs D-f, P-f, P-s, and D are the four designs presented in Figure~\ref{fig:cloak_1}, and designs (1) to (16) are the sixteen designs illustrated in Figure~\ref{fig:cloak_2}.}
    \label{fig:diff_b}
\end{figure*}

\begin{figure*}[t]
    \centering
    \includegraphics[scale=0.3]{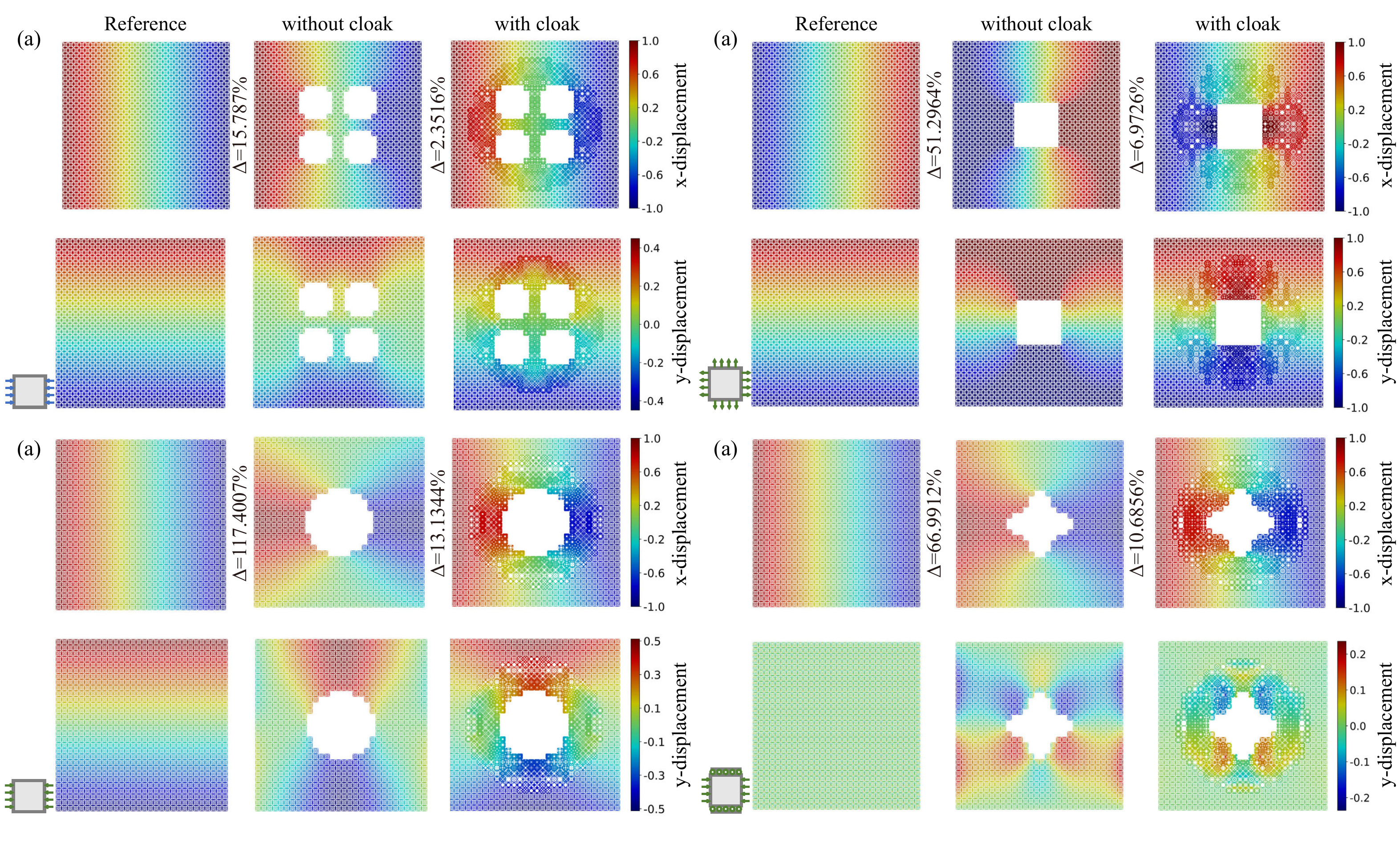}
    \caption{Mechanical cloak design results under different boundary conditions and void shapes using microstructure dataset (1). (a) The results of cloak structure for four circular voids under displacement-free boundary condition (b) The results of cloak structure for circular void under pressure-free boundary condition (c) The results of cloak structure for square void under dilating boundary condition (d) The results of cloak structure for star void under pressure-sliding boundary condition.}
    \label{fig:cloak_1}
\end{figure*}

\begin{figure*}[t]
    \centering
    \includegraphics[scale=0.22]{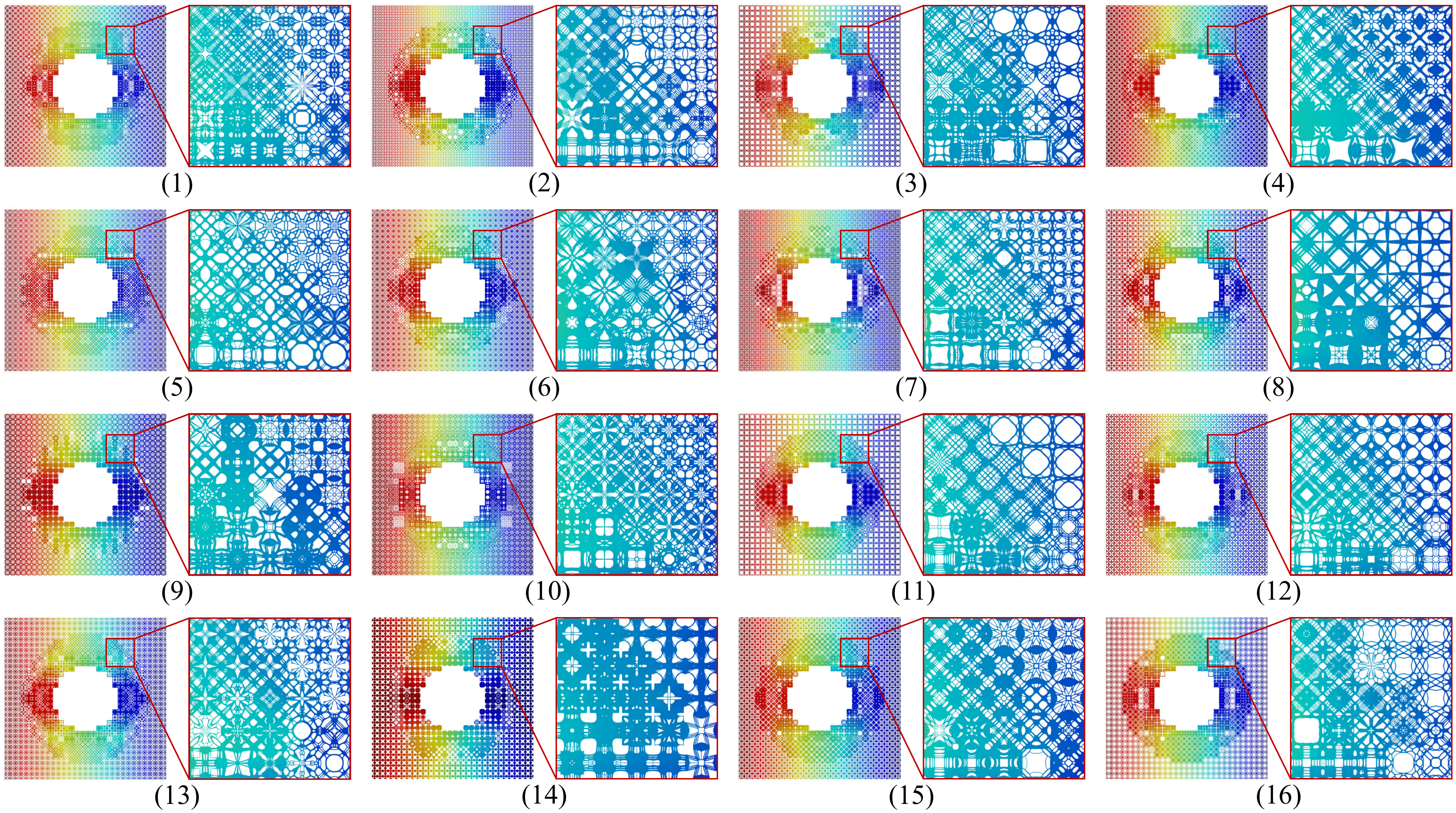}
    \caption{Cloak design results under displacement-free boundaries and circular void using 16 different datasets. The red box shows a partial enlargement of the cloak.}
    \label{fig:cloak_2}
\end{figure*}

\begin{figure*}[!t]
    \centering
    \includegraphics[scale=0.22]{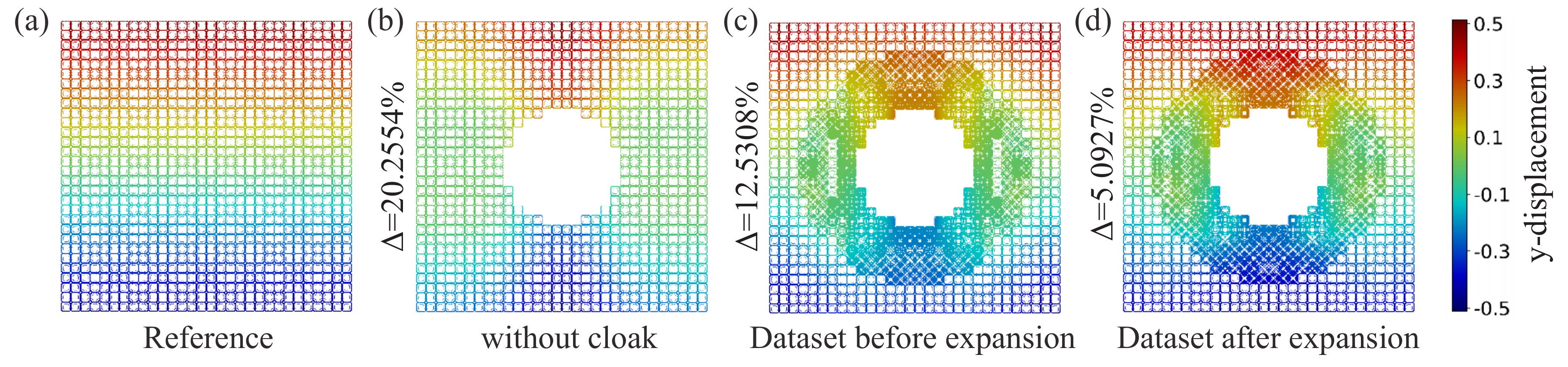}
    \caption{Mechanical cloaks designed using different datasets, where colors indicate the y-direction displacement. (a) The original structure, (b) Structure with void, (c) Structure with a cloak designed using the dataset before expansion, (d) Structure with a cloak designed using the final dataset.}
    \label{fig:multi3}
\end{figure*}

\begin{figure*}[t]
    \centering
    \includegraphics[scale=0.38]{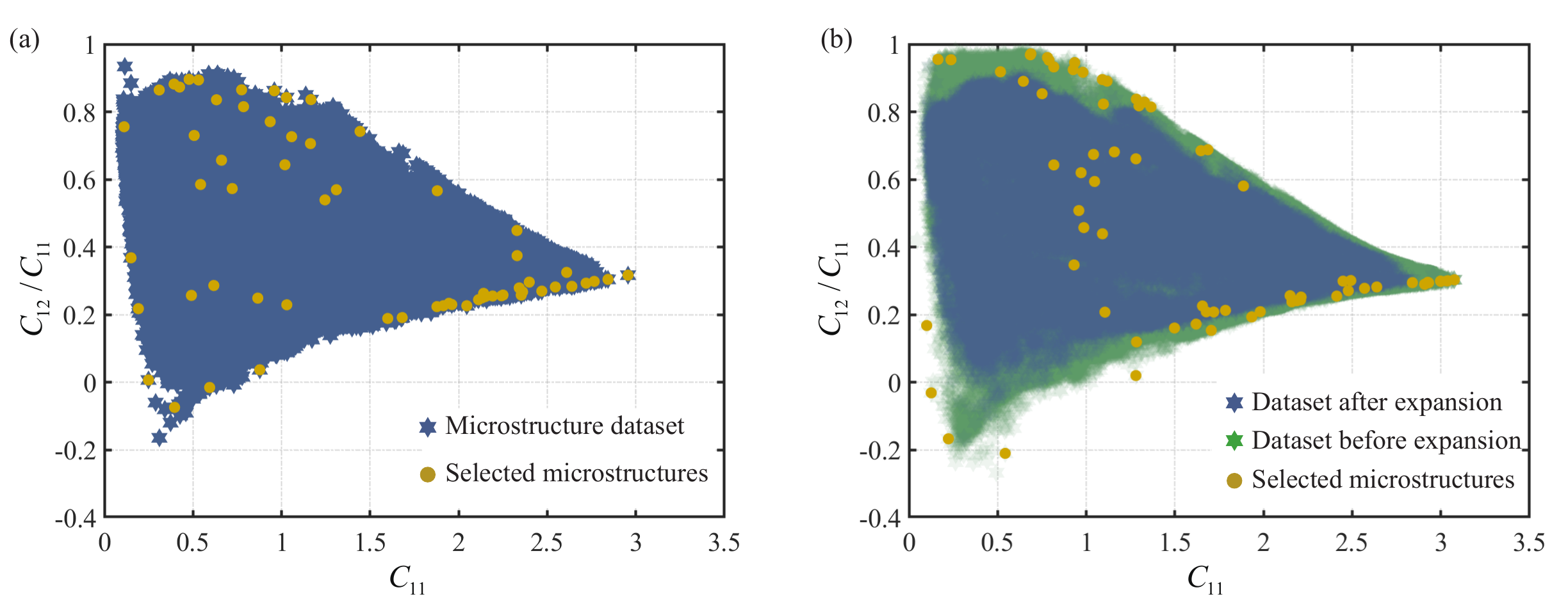}
    \caption{Range of properties for different microstructure datasets and the properties of selected microstructures for mechanical cloak design: (a) dataset before expansion (b) dataset after expansion. }
    \label{fig:multi1}
\end{figure*}

\begin{figure*}[t]
    \centering
    \includegraphics[scale=0.23]{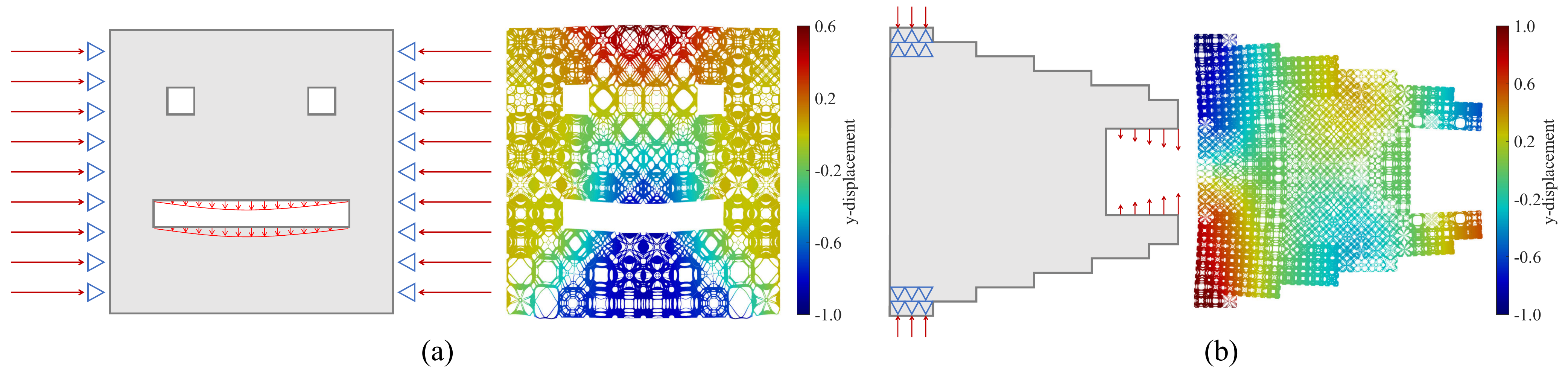}
    \caption{The target settings and displacement simulation of optimized results for (a) smiley face filled with $10\times 10$ macro unit cells and (b) gripper filled with $20\times 20$ macro unit cells. }
    \label{fig:face}
\end{figure*}
\subsection{Multiscale topology optimization}
\label{sec:multiscale}

%To demonstrate the approach, we apply the generated boundary-identical microstructure datasets to design two applications of multiscale microstructure systems, mechanical cloak design, and customized displacement design.  

\subsubsection{Mechanical cloak design}
For mechanical cloak design, the reference structures (without voids and cloak) are composed of $30\times 30$ periodically tessellated base cells chosen in the microstructure dataset. The mechanical cloak is defined in a circle of radius 12 at the center of the design domain.

As shown in Figure~\ref{fig:diff_b} (a), four different boundaries are employed. Notably, the blue arrow signifies that the designated displacement represents the boundary condition $u_{bc}$, while the green arrow indicates that the specified force serves as the boundary condition $p_{bc}$. Figure~\ref{fig:diff_b} (b) illustrates the relative displacement difference before and after the incorporation of the mechanical cloak. For a quantitative measure of the distortion of the displacement field, the relative displacement difference $\Delta$ is formulated as 

\begin{equation}
    \Delta = \frac{\sqrt{\sum_{\Omega_s} (\bm{u}_o-\bm{u}_t)^2}}{\sqrt{\sum_{\Omega_s} (\bm{u}_t)^2}},
    \label{eq:error}
\end{equation}
where $\bm{u}_o$ and $\bm{u}_t$ are the node displacement of mechanical cloak and reference structures. The difference, as indicated by the red bar with the cloak applied, is significantly lower compared to the blue bar without the cloak, especially for the second boundary (13.13\% vs. 117.40\%). 

Figure~\ref{fig:cloak_1} depicts the outcomes of the cloak design for four distinct boundary conditions using the microstructure dataset (1) shown in Figure~\ref{fig:one-result}. In Figure~\ref{fig:cloak_1} (b) and (d), it is evident that under the pressure-free and dilating boundary conditions, the displacements in the $x$ and $y$ directions in the results without the cloak significantly deviate from those of the reference structures. The addition of the cloak structure notably reduces the displacement gap. The dataset generated by our method shown in Figure~\ref{fig:one-result} is capable of addressing complex deformation scenarios while maintaining full boundary connectivity, thus reducing the error of the above four examples to less than 13.2\%    .%resulting in an error reduction of less than 13.2\% for the aforementioned four examples.

To validate the design capabilities of mechanical cloaks for additional datasets, we conducted tests on 16 datasets shown in Figure~\ref{fig:datasets} under displacement-free boundary conditions, each featuring a hollow circular hole in the center. The results and statistics are shown in Figure~\ref{fig:cloak_2} and Figure~\ref{fig:diff_b}, respectively. 
The results obtained from the analysis of 16 datasets revealed a maximum difference of 5.6\%. The generated datasets encompass a broad spectrum and has the potential to facilitate the design of mechanical cloaks.

%\paragraph{Advantages of generating new microstructures for multi-scale optimization}

To evaluate the impact of dataset iterative expansion on multiscale design outcomes, we conduct the following experiment: We select the 15-th dataset, both before and after expansion,  to perform identical mechanical cloak designs. Before filling in the mechanical cloaks, the relative displacement difference was $\Delta = 20.2554\%$. Through multiscale topology optimization, two mechanical cloak designs as shown in Figure~\ref{fig:multi3} are obtained. The design using the 15-th dataset before expansion obtains a relative displacement difference of $\Delta = 12.5308\%$, while the expansion dataset yields a difference of $\Delta = 5.0927\%$.

The positions of elastic tensors for the microstructures used in the mechanical cloaks are shown in Figure~\ref{fig:multi1}. Most of microstructures are located at the periphery. The cloak designed with the expanded dataset outperforms that of the pre-expansion dataset, likely due to the latter's narrower property coverage, which limited further performance improvements. This highlights the benefit of broader property coverage for this problem and confirms that dataset coverage significantly impacts design performance.

\subsubsection{Customized displacement design}
For the second application, two examples (a face and a gripper) are designed to show the effect of customized displacement design. 
For the multiscale structure design of the face, we opt for a combination of $10\times10$ unit cells and utilize the microstructure dataset presented in Figure~\ref{fig:one-result} for the selection and assembly of microstructures. 
When pressure is applied to both sides of the face, the mouth will exhibit a downward curve, creating the appearance of a smile expression. Likewise, in the gripper design, a grid of $20\times20$ unit cells is chosen, and the microstructure dataset (5) in Figure~\ref{fig:datasets} served as the microstructure library for optimal design. When the gripper is compressed vertically on the left side, the structure on the right side contracts towards the central axis. Simulation results indicate that the optimized outcomes align closely with the specified displacement shown in Figure~\ref{fig:face}. Applying the calculation formula in Eq.~\eqref{eq:error}, the relative differences for the face and gripper are 12.19\% and 8.70\%, respectively. 

\begin{figure*}[htbp]
    \centering
    \includegraphics[scale=0.5]{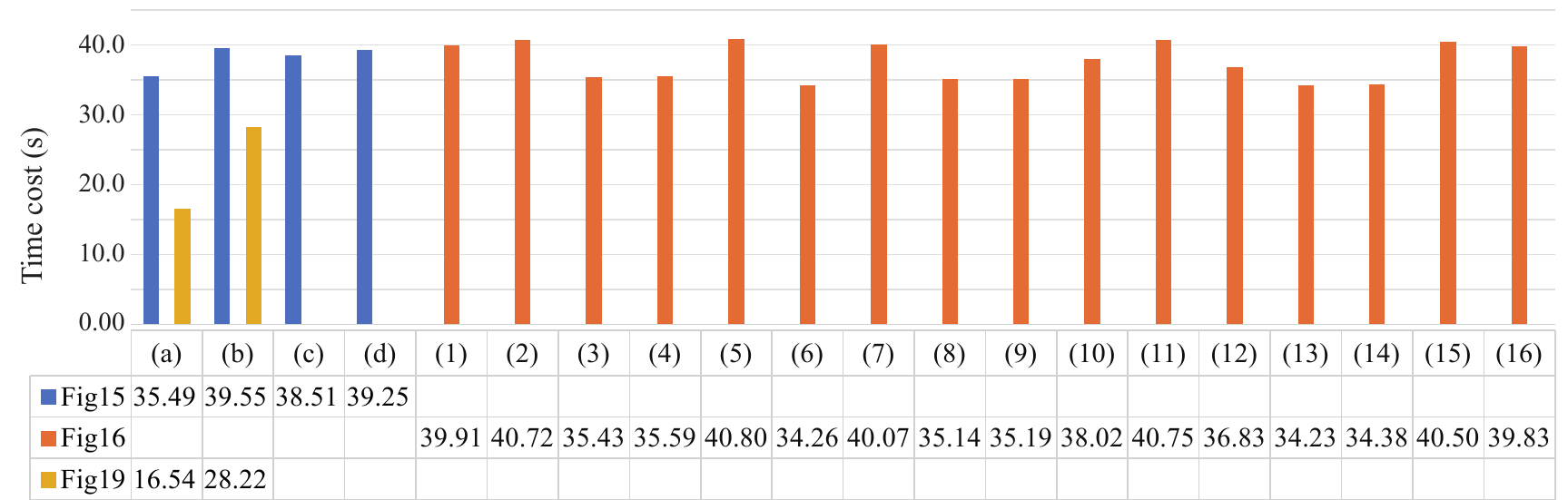}
    \caption{The statistical results for the optimization time of %mechanical cloaks
    multiscale designs in Figure~\ref{fig:cloak_1},~\ref{fig:cloak_2},~\ref{fig:face} include optimizing the elastic modulus of macroscopic elements and searching for the microstructure corresponding to the target modulus.}
    \label{fig:time}
\end{figure*}

\subsubsection{Optimization efficiency}
The present study underscores a notable enhancement in the algorithm's efficiency, evidenced by a substantial reduction in the computation time. 
The computational time for multiscale topology optimization is primarily allocated to two tasks: (1) optimizing the elastic modulus of macroscopic elements and (2) searching for the microstructure corresponding to the target modulus. 
Specifically, the algorithm has achieved a remarkable improvement, all tested examples are completed in one minute. The specific time is shown in Figure~\ref{fig:time}. 
%reducing the original calculation time from 10 hours~\cite{wang2020deep} to less than 1 minute. 
%\tb{In Figure~\ref{fig:time}, we present the calculation time for each cloak, which includes both macroscopic optimization and the process of searching the database for suitable microstructures to fill the design.} 
%This advancement signifies a notable 600-fold increase in efficiency. 
In the data-driven approach~\cite{wang2020deep} for computing top-down multiscale problems, conducting a single-cell search to ensure connectivity imposes a considerable time burden.% In the computation of top-down multiscale problems by data-driven approach~\cite{wang2020deep}, it is necessary to conduct a single cell search to ensure connectivity imposes a considerable time burden. 

\section{Conclusion and Future Work} 

This paper introduces 16 large-range cubic symmetric microstructure datasets with identical boundaries and a wide range of elastic moduli generated through a deep generative model. 
The self-conditioning diffusion model derives microstructures under the specified boundary and the pre-defined homogenized elastic tensor. 
Initially, an inverse homogenization approach is utilized to generate the optimized dataset. 
Subsequently, the microstructure boundaries of the dataset are clustered to categorize it into 16 cases. 
Then, through the expansion process, a series of datasets with a broad range of elastic moduli and identical boundaries are generated. 
We showcase the data generation capabilities of this generative model by designing two multiscale structures: mechanical cloaks and structures with specified displacements. 
By employing numerous examples featuring varying numbers and shapes of voids and diverse boundaries, we illustrate that our method excels in multiscale systems. 
This superiority is validated through numerical simulations and computational efficiency. %The proposed datasets are characterized by consistent boundaries, thereby ensuring the natural resolution of the connectivity problem. 
Besides, this results in a substantial reduction in calculation time.

Three-dimensional microstructures find more widespread applications, and we intend to explore the extension of the current results to three-dimensional microstructures in future work. This extension poses two potential challenges: 
(1) Generating voxel-based 3D microstructure libraries demands significant computational time and memory resources. Balancing generation efficiency with effectiveness becomes a critical consideration.
(2) Classifying boundaries for 3D microstructures is notably more complex than for their 2D counterparts, which presents a challenging task in determining reasonable boundary categories for 3D microstructures.

%\vspace{-0.5cm}

\section*{Acknowledgement}
This work is supported by the National Natural Science Foundation of China (No. 62402467, No. 62025207, No. 61972368), the National Key R\&D Program of China (No. 2024YFA1016300),  the Youth Innovation Key Research Funds for the Central Universities, China (No. YD0010002010), the Strategic Priority Research Program of the Chinese Academy of Sciences (No. XDB0640000). 

% %\vspace{-0.5cm}
\section*{Declarations}
% \paragraph{Conflict of interest} 
The authors declare that they have no conflict of interest. 

% \paragraph{Replication of results} Code for this paper is available at \url{https://github.com/Schnabel-8/BDR-diffusion}, and the datasets are publicly available at \href{https://rec.ustc.edu.cn/share/b9e072b0-306a-11ef-91cb-dfb6325d6cc5}{https://rec.ustc.edu.cn/share/dataset}%\url{https://rec.ustc.edu.cn/share/b9e072b0-306a-11ef-91cb-dfb6325d6cc5}.  %Code for this paper is at ~\url{https://github.com/quanyuchen2000/OPENTM}.

\bibliography{mybibfile}

\appendix
\setcounter{figure}{0}
\renewcommand{\thefigure}{A\arabic{figure}}

\setcounter{table}{0}
\renewcommand{\thetable}{A\arabic{table}}
%\section*{Appendix}

%\subsection*{\tr{...}}
\begin{figure*}[!t]
    \centering
    \includegraphics[scale=0.085]{hsbound_1.pdf}
    \caption{The bulk modulus of the microstructures in the 16 datasets. The color bar indicates the proportion of the microstructure's bulk modulus to the corresponding HS-bound. The red line represents HS-bound and the red star represents the selected microstructures that reach 98\% HS-bound. }
    \label{fig:hsbound}
\end{figure*}

\section{Analysis of Bulk Modulus for the Datasets}

We analyze the bulk modulus of the 16 datasets constructed in this study and compared them with the HS-bound. The results of this comparison are presented in Figure ~\ref{fig:hsbound}.

% 消融实验结果
% w/o self-conditioning和baseline相比平均降低5.16，最大降低11.20
% 后三个和baseline相比平均降低2.21, 3.98, 4.26
% 最大降低8.80, 14.70, 16.00

\section{Additional results}

\paragraph{Ablation study results} Study on self-conditioning and elastic tensor embedding includes a base model and four variations, each omitting either self-conditioning or elastic tensor embedding in low, mid, or high-level layers. The models are evaluated under conditions of specified target elastic tensors and 16 boundaries. Table \ref{tab:performance1} shows the $R^2$-scores between the generated and target elastic tensors. Table \ref{tab:performance3} presents the degree of match for the boundaries of the generated microstructures against the target boundaries.
\begin{table}[ht]
\centering
\caption{Generation accuracy evaluated by the $R^2$-score of different models under 16 boundaries. 
}
\label{tab:performance1}
\begin{tabularx}{\textwidth}{@{}l*{8}{>{\centering\arraybackslash}X}@{}}
\toprule
%\multirow{2}{*}{\textbf{\makecell[c]{Method}}} & \multicolumn{8}{c}{\textbf{Datasets}} \\
%\cmidrule(l){2-9}
 Methods & 1 & 2 & 3 & 4 & 5 & 6 & 7 & 8 \\
\midrule
baseline & 0.958 & 0.950 & 0.962 & 0.957 & 0.968 & 0.971 & 0.985 & 0.977 \\
w/o self-conditioning & 0.870 & 0.924 & 0.939 & 0.920 & 0.925 & 0.921 & 0.941 & 0.936 \\
\makecell[l]{w/o low-level layers \\ elastic tensor embedding} & 0.896 & 0.917 & 0.960 & 0.944 & 0.951 & 0.970 & 0.985 & 0.974 \\
\makecell[l]{w/o middle layers \\ elastic tensor embedding } & 0.906 & 0.875 & 0.948 & 0.956 & 0.922 & 0.969 & 0.918 & 0.973 \\
\makecell[l]{w/o high-level layers \\ elastic tensor embedding} & 0.840 & 0.881 & 0.961 & 0.955 & 0.911 & 0.934 & 0.984 & 0.970 \\
\bottomrule
 Methods  & 9 & 10 & 11 & 12 & 13 & 14 & 15 & 16 \\
\midrule
baseline & 0.963 & 0.967 & 0.971 & 0.941 & 0.950 & 0.956 & 0.994 & 0.975 \\
w/o self-conditioning & 0.914 & 0.932 & 0.859 & 0.899 & 0.847 & 0.914 & 0.935 & 0.943 \\
\makecell[l]{w/o low-level layers \\ elastic tensor embedding} & 0.963 & 0.950 & 0.883 & 0.936 & 0.902 & 0.948 & 0.975 & 0.937 \\
\makecell[l]{w/o middle layers \\ elastic tensor embedding} & 0.923 & 0.939 & 0.883 & 0.934 & 0.803 & 0.914 & 0.971 & 0.974 \\
\makecell[l]{w/o high-level layers \\ elastic tensor embedding} & 0.957 & 0.914 & 0.811 & 0.924 & 0.859 & 0.919 & 0.970 & 0.973 \\
\bottomrule
\end{tabularx}
\end{table}
% self-conditioning平均降低2.55，最大降低7.00
% 后三行平均降低1.49, 1.63, 1.87
% 最大降低 3.80, 5.80, 5.10
\begin{table}[ht]
\centering
\caption{
Generation accuracy evaluated by the DoM value of different models under 16 boundaries 
}
\label{tab:performance3}
\begin{tabularx}{\textwidth}{@{}l*{8}{>{\centering\arraybackslash}X}@{}}
\toprule
% \multirow{2}{*}{\textbf{\makecell[c]{Method}}} & \multicolumn{8}{c}{\textbf{Datasets}} \\
% \cmidrule(l){2-9}
Methods & 1 & 2 & 3 & 4 & 5 & 6 & 7 & 8 \\
\midrule
baseline & 0.985 & 0.993 & 0.990 & 0.987 & 0.991 & 0.989 & 0.997 & 0.994 \\
w/o self-conditioning & 0.962 & 0.976 & 0.974 & 0.978 & 0.973 & 0.987 & 0.989 & 0.985 \\
\makecell[l]{w/o low-level layers \\ elastic tensor embedding} & 0.951 & 0.989 & 0.986 & 0.980 & 0.968 & 0.982 & 0.985 & 0.957 \\
\makecell[l]{w/o middle layers \\ elastic tensor embedding} & 0.967 & 0.981 & 0.985 & 0.983 & 0.974 & 0.989 & 0.976 & 0.992 \\
\makecell[l]{w/o high-level layers \\ elastic tensor embedding} & 0.939 & 0.975 & 0.983 & 0.971 & 0.986 & 0.978 & 0.984 & 0.990 \\
\bottomrule
Methods & 9 & 10 & 11 & 12 & 13 & 14 & 15 & 16 \\
\midrule
baseline & 0.992 & 0.987 & 0.991 & 0.989 & 0.995 & 0.994 & 0.998 & 0.996 \\
w/o self-conditioning & 0.953 & 0.975 & 0.921 & 0.934 & 0.962 & 0.926 & 0.988 & 0.977 \\
\makecell[l]{w/o low-level layers \\ elastic tensor embedding} & 0.986 & 0.989 & 0.953 & 0.971 & 0.985 & 0.980 & 0.983 & 0.984 \\
\makecell[l]{w/o middle layers \\ elastic tensor embedding} & 0.934 & 0.988 & 0.963 & 0.982 & 0.951 & 0.987 & 0.975 & 0.980 \\
\makecell[l]{w/o high-level layers \\ elastic tensor embedding} & 0.979 & 0.965 & 0.942 & 0.980 & 0.944 & 0.988 & 0.984 & 0.981 \\
\bottomrule
\end{tabularx}
\end{table}

%\section{}

\paragraph{Elastic tensor coverage ratios over expansion iterations} Table ~\ref{tab:range1} illustrates changes in coverage ratios for the 16 microstructure datasets with identical boundaries during iterative expansion processes using active learning method. It shows improvements in overall coverage expansion. As shown, the expansion iterations result in the effective expansion of the elastic tensor coverage within the microstructure datasets.

\begin{table}[h]
\centering
\caption{The statistics of coverage ratios for datasets 1-16 from iterations 0 to 3.}
\label{tab:range1}
\begin{tabularx}{\textwidth}{@{}l*{8}{>{\centering\arraybackslash}X}@{}}
\toprule
% \multirow{2}{*}{\textbf{\centering Iterations}} & \multicolumn{8}{c}{\textbf{Datasets}} \\
% \cmidrule(l){2-9}
Iterations & 1 & 2 & 3 & 4 & 5 & 6 & 7 & 8 \\
\midrule
~~~~~~~0 & 0.00785 & 0.01378 & 0.00900 & 0.00848 & 0.00774 & 0.00689 & 0.00778 & 0.00821 \\
~~~~~~~1 & 0.00845 & 0.01501 & 0.00945 & 0.00909 & 0.00844 & 0.00778 & 0.00863 & 0.00886 \\
~~~~~~~2 & 0.00884 & 0.01594 & 0.00983 & 0.00990 & 0.00903 & 0.00813 & 0.00923 & 0.00941 \\
~~~~~~~3 & 0.00915 & 0.01633 & 0.01035 & 0.01018 & 0.00941 & 0.00847 & 0.00942 & 0.00968 \\
\bottomrule
% \end{tabularx}
% \end{table}
% \begin{table}[htbp]
% \centering
% \caption{The coverage of propertys for datasets 9 through 16 during iterations 0 through 3.}
% \label{tab:range2}
% \begin{tabularx}{\textwidth}{@{}l*{8}{>{\centering\arraybackslash}X}@{}}
%\toprule
% \multirow{2}{*}{\textbf{\centering Iterations}} & \multicolumn{8}{c}{\textbf{Datasets}} \\
% \cmidrule(l){2-9}
Iterations& 9 & 10 & 11 & 12 & 13 & 14 & 15 & 16 \\
\midrule
~~~~~~~0 & 0.00858 & 0.00519 & 0.00827 & 0.00765 & 0.01685 & 0.01269 & 0.01601 & 0.01304 \\
~~~~~~~1 & 0.00961 & 0.00591 & 0.00923 & 0.00825 & 0.01824 & 0.01416 & 0.01796 & 0.01471 \\
~~~~~~~2 & 0.01015 & 0.00636 & 0.00969 & 0.00920 & 0.01916 & 0.01491 & 0.01968 & 0.01560 \\
~~~~~~~3 & 0.01059 & 0.00662 & 0.01032 & 0.00986 & 0.01990 & 0.01554 & 0.02006 & 0.01633 \\
\bottomrule
\end{tabularx}
\end{table}

\end{document}